\documentclass[%
reprint,
superscriptaddress,
groupedaddress,
nofootinbib,
nobibnotes,
 amsmath,amssymb,
 aps,
prl,
]{revtex4-2}
\usepackage[dvipsnames]{xcolor}
\usepackage{graphicx}
\usepackage{dcolumn}
\usepackage{bm}
\usepackage[breaklinks=true,colorlinks,citecolor=blue,linkcolor=red,urlcolor=blue]{hyperref}

\usepackage{braket}
\usepackage[utf8]{inputenc} 
\usepackage{mathtools}
\usepackage[bottom]{footmisc}
\usepackage{hyperref}
\usepackage{amsfonts}
\usepackage{amsmath}
\usepackage{slashed}
\usepackage{amsthm}

\usepackage{tikz}

\begin{document}



\title{Geometry-Driven Moiré Engineering in Twisted Bilayers of High-Pseudospin Fermions}

\author{Yi-Chun Hung$\,^{\hyperlink{equal}{*}}$}

\author{Xiaoting Zhou$\,^{\hyperlink{equal}{*},\hyperlink{email1}{\dagger}}$}

\author{Arun Bansil$\,^{\hyperlink{email2}{\ddagger}}$}

\affiliation{Department of Physics,\;Northeastern\;University,\;Boston,\;Massachusetts\;02115,\;USA}
\affiliation{Quantum Materials and Sensing Institute,\;Northeastern University,\;Burlington,\;Massachusetts\;01803,\;USA}


\begin{abstract}
Moiré engineering offers new pathways for manipulating emergent states in twisted layered materials and lattice-mismatched heterostructures. With the key role of the geometry of the underlying lattice in mind, here we introduce the \emph{\textbf{watermill lattice}}, a two-dimensional structure with low-energy states characterized by massless pseudospin-3/2 fermions with high winding numbers. Its twisted bilayer is shown to exhibit magic angles, where four isolated flat bands emerge around the Fermi level, featuring elevated Wilson-loop windings and enhanced quantum geometric effects, such as an increase in the ratio of the Berezinskii-Kosterlitz-Thouless (BKT) transition temperature to the mean-field critical temperature under a weak Bardeen-Cooper-Schrieffer (BCS) pairing. We discuss how the watermill lattice could be realized in the MXene and group-IV materials. Our study highlights the potential of exploiting lattice geometry in moiré engineering to uncover novel quantum phenomena and tailor emergent electronic properties in materials.
\end{abstract}


\maketitle
\renewcommand{\thefootnote}{\fnsymbol{footnote}}
\footnotetext[1]{\hypertarget{equal}{These authors contributed equally.}}
\footnotetext[2]{\hypertarget{email1}{Contact author: \href{mailto:x.zhou@northeastern.edu}{x.zhou@northeastern.edu}}}
\footnotetext[3]{\hypertarget{email2}{Contact author: \href{mailto:ar.bansil@northeastern.edu}{ar.bansil@northeastern.edu}}}
\par Moiré engineering is drawing intense interest as an innovative technique for manipulating low-dimensional electronic structures by leveraging emergent potentials arising from moiré patterns in twisted-layered materials and lattice-mismatched heterostructures \cite{Andrei2021, doi:10.1126/sciadv.1601459, WINTTERLIN20091841, Andrei2020}. A remarkable consequence  is the emergence of flat electronic bands with substantial quantum metric, which allows the exploration of exotic quantum phenomena, including strongly correlated physics and effects of quantum geometry on physical properties, such as superfluid weights \cite{PhysRevLett.124.167002, PhysRevLett.128.087002, PhysRevB.95.024515, Peotta2015, PhysRevLett.117.045303} and light-matter interactions \cite{doi:10.1126/science.adg0014, PhysRevB.104.064306, tai2023quantumgeometriclightmattercouplingcorrelated, PhysRevResearch.4.013164}.

\par The physics of twisted bilayer systems, such as twisted bilayer graphene (TBG) \cite{Andrei2020, doi:10.1073/pnas.1108174108, PhysRevLett.122.106405, PhysRevB.99.155415, PhysRevLett.123.036401, PhysRevLett.124.167002, Cao2018, Cao2018_2, Oh2021, doi:10.1126/science.aav1910, PhysRevX.8.041041, PhysRevX.8.031089, PhysRevB.104.115404} and transition metal dichalcogenides (TMDs) \cite{PhysRevLett.122.086402, doi:10.1073/pnas.2021826118, Devakul2021, Wang2020, Guo2025, Cai2023, Park2023, PhysRevB.108.085117, PhysRevX.13.031037, Jin2019, Yuan2020, Shimazaki2020, Jiang2021, Wilson2021}, is driven by the underlying lattice structures and twist angles and the associated intricate moiré patterns. At specific "magic angles," flat bands are generated, which enable the exploration of exotic quantum phenomena, such as unconventional superconductivity \cite{Balents2020, Cao2018, Cao2018_2, Oh2021, doi:10.1126/science.aav1910, PhysRevX.8.041041, PhysRevX.8.031089, Wang2020, Guo2025}, fractional quantum anomalous Hall effects \cite{Cai2023, Park2023, PhysRevB.108.085117, PhysRevX.13.031037}, excitonic physics \cite{Jin2019, Yuan2020, Shimazaki2020, Jiang2021, Wilson2021, Tran2019}, and complex magnetism \cite{PhysRevLett.128.217202, PhysRevB.98.245103, PhysRevLett.122.246402, PhysRevB.102.201104, PhysRevResearch.2.033087, PhysRevB.110.085135, AkifKeskiner2024}. Moiré physics in twisted-bilayer anisotropic lattices can induce stripes and quasi-one-dimensional behaviors, leading to density waves and Luttinger liquid phases \cite{D1NR07736H, PhysRevLett.133.246501, PhysRevB.108.L201120, PhysRevB.96.195406, PhysRevB.104.165130, pub.1170483970, PhysRevB.108.L121409, chang2024twodimensionalspinhelixmagnoninduced}.

\par Relatively simple minimalistic models have often sparked breakthroughs in condensed matter physics by stripping away complexity to isolate essential physics and uncover new insights into the nature of quantum matter. Two-dimensional (2D) lattice models, such as the kagome \cite{10.1143/ptp/6.3.306, PhysRevB.99.125131, PhysRevB.101.045131}, dice \cite{PhysRevB.106.155417, PhysRevB.108.075166, PhysRevB.108.075167, PhysRevB.99.155124, PhysRevB.107.035421}, Lieb \cite{PhysRevLett.62.1201, PhysRevB.99.125131, PhysRevB.101.045131} and checkerboard lattices \cite{PhysRevB.78.245122, PhysRevLett.103.046811} showcase unique lattice geometries that give rise to distinctive quantum geometric phenomena. Effects of quantum geometry become especially enhanced in moiré structures constructed from special 2D lattices. 

A recent example is provided by high pseudospin $1$ low-energy states (LESs) in twisted bilayers of Dice lattices which support a tunable number of isolated flat bands near the Fermi level \cite{PhysRevLett.133.236401, PhysRevB.109.155159}. This naturally prompts the question: What unique behaviors might arise in lattices with LESs characterized by higher pseudospins, such as $3/2$? Accordingly, here we introduce the novel 2D \emph{watermill lattice}, which hosts pseudospin $3/2$ LESs, and the moiré physics of twisted bilayer watermill lattice. Experimental realizations of the watermill lattice in real materials are also discussed. 

\begin{figure}[h]
\centering
\includegraphics[width=\linewidth]{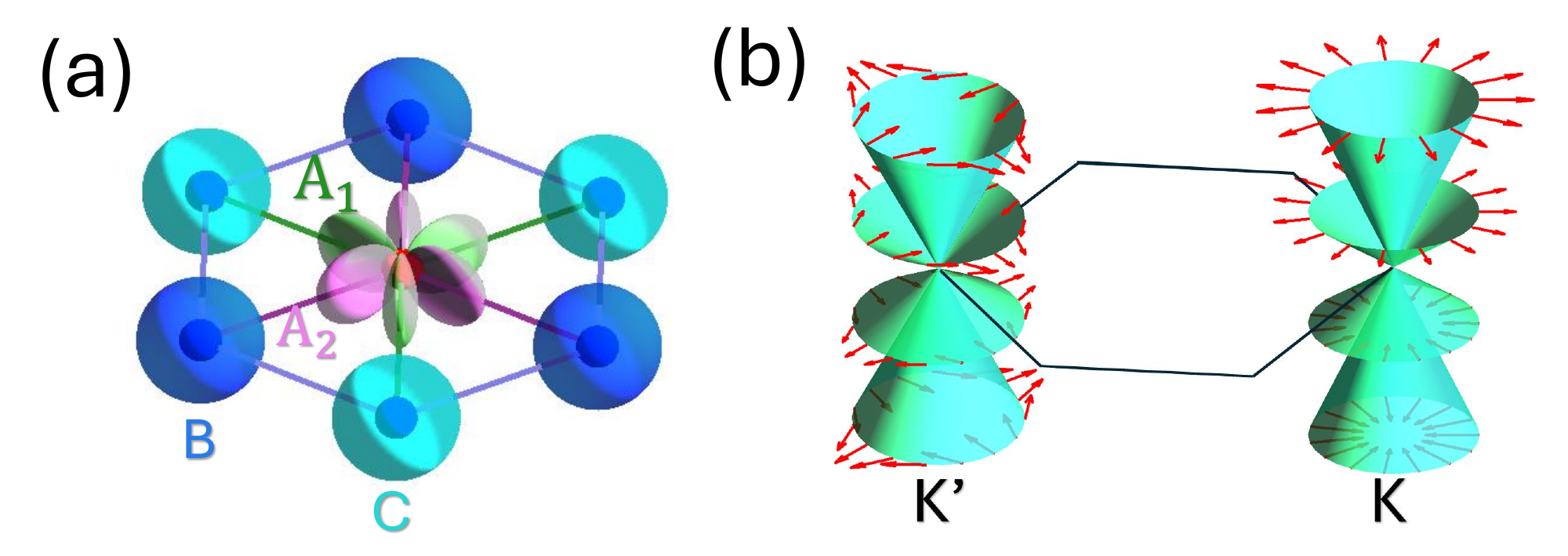}
\caption{ (a) A schematic diagram of the watermill lattice. Nearest-neighbor hopping between orbitals $B$ and $C$ are marked in blue, and between orbitals $A_1 (A_2)$ and orbital $B (C)$ are marked in green (magenta). (b) Band structure of the watermill lattice near the $K,K'$-valley.}
\label{fig:01}
\end{figure}

\paragraph*{The watermill lattice---}Inspired by the Dice lattice, we introduce the watermill lattice. It consists of a honeycomb lattice augmented by a central atom, labeled $A$, which with two orbitals, $A_1$ and $A_2$. The original honeycomb lattice sites retain their single orbitals, B and C. We focus on the nearest-neighbor hopping interactions within this lattice, with hopping between site A and its neighboring B (C) sites mediated by orbitals $A_1$ ($A_2$), as depicted in FIG.~\ref{fig:01}(a). 

\par Note that the atoms at sites $B$ and $C$ are identical, whereas the central atom at site $A$ is distinct. The hopping interactions between pairs of sites $A$-$B(C)$ involve orbitals $A_1 (A_2)$ and $B (C)$, respectively. Due to the six-fold rotational symmetry $C_{6z}$, the hopping strengths between $A$-$B$ and $A$-$C$, are equal, denoted as $t'$, but differ from those between $B$ and $C$, denoted as $t$. We set $t' = \frac{\sqrt{3}}{2}t$. Consequently, in the basis of  $\Psi = {(A_1, B, C, A_2)}^T$, the Bloch Hamiltonian can be written as: 
\begin{equation}\label{eq:01}
    H_{\text{WM}}(\vec{k}) = t\begin{pmatrix} 0 & \frac{\sqrt{3}}{2}h(\vec{k})  & 0 & 0 \\ \frac{\sqrt{3}}{2}h(\vec{k}) ^* & 0 & h(\vec{k})  & 0 \\ 0 & h(\vec{k}) * & 0 & \frac{\sqrt{3}}{2}h(\vec{k})  \\ 0 & 0 & \frac{\sqrt{3}}{2}h(\vec{k}) ^* & 0 \end{pmatrix},
\end{equation}
where $h(\vec{k}) =1+2\cos(\frac{\sqrt{3}k_x}{2})e^{-i\frac{3k_y}{2}}$.

\par The watermill lattice, characterized by the hopping parameters outlined above, exhibits time-reversal symmetry (TRS) $\mathcal{T}=\mathcal{K}$, sixfold rotational symmetry $C_{6z}$, chiral symmetry $\Gamma$, particle-hole symmetry $\Xi$, and the combined symmetry $C_{2z}\mathcal{T}$.  These symmetries are tied to the lattice structure. For the Hamiltonian in Eq.~\eqref{eq:01}, these symmetries are represented by $\Gamma=\sigma_0\otimes\sigma_3$,  $\Xi=\sigma_0\otimes\sigma_3\mathcal{K}$, and $C_{2z}\mathcal{T} = \sigma_1\otimes\sigma_1\mathcal{K}$, where $\mathcal{K}$ is the complex conjugation operator, $\sigma_{1,2,3}$ are Pauli matrices, and $\sigma_0$ is the identity matrix.

\par The watermill lattice exhibits valley structures with four-fold degenerate cones at the $K$ and $K'$ points. The low-energy effective Hamiltonian around the valleys in the continuum limit is:
\begin{equation}\label{eq:02}
    H_0(\vec{q}) = \frac{3t}{2}(q_x S_1 +\xi q_y S_2).
\end{equation}
Here, $\xi=\pm$ designates the valley index, where $\xi= +$ refers to the $K$ valley, and $\xi= -$ to the $K'$ valley. $\vec{q}=\vec{k}-K^{(\prime)}$, and $S_{i}$ denote the spin matrices for a spin-3/2 fermion \cite{Griffiths_Schroeter_2018}. Equation \eqref{eq:02} describes a four-fold degenerate pseudospin-3/2 quasiparticle excitation, protected by the sublattice symmetry at $K$ and $K'$ points in the Brillouin zone (BZ). Furthermore, each band exhibits a nontrivial pseudo-spin texture with winding number $w = \xi$, as illustrated in FIG.~\ref{fig:01}(b).

\begin{figure}[h]
\centering
\includegraphics[width=\linewidth]{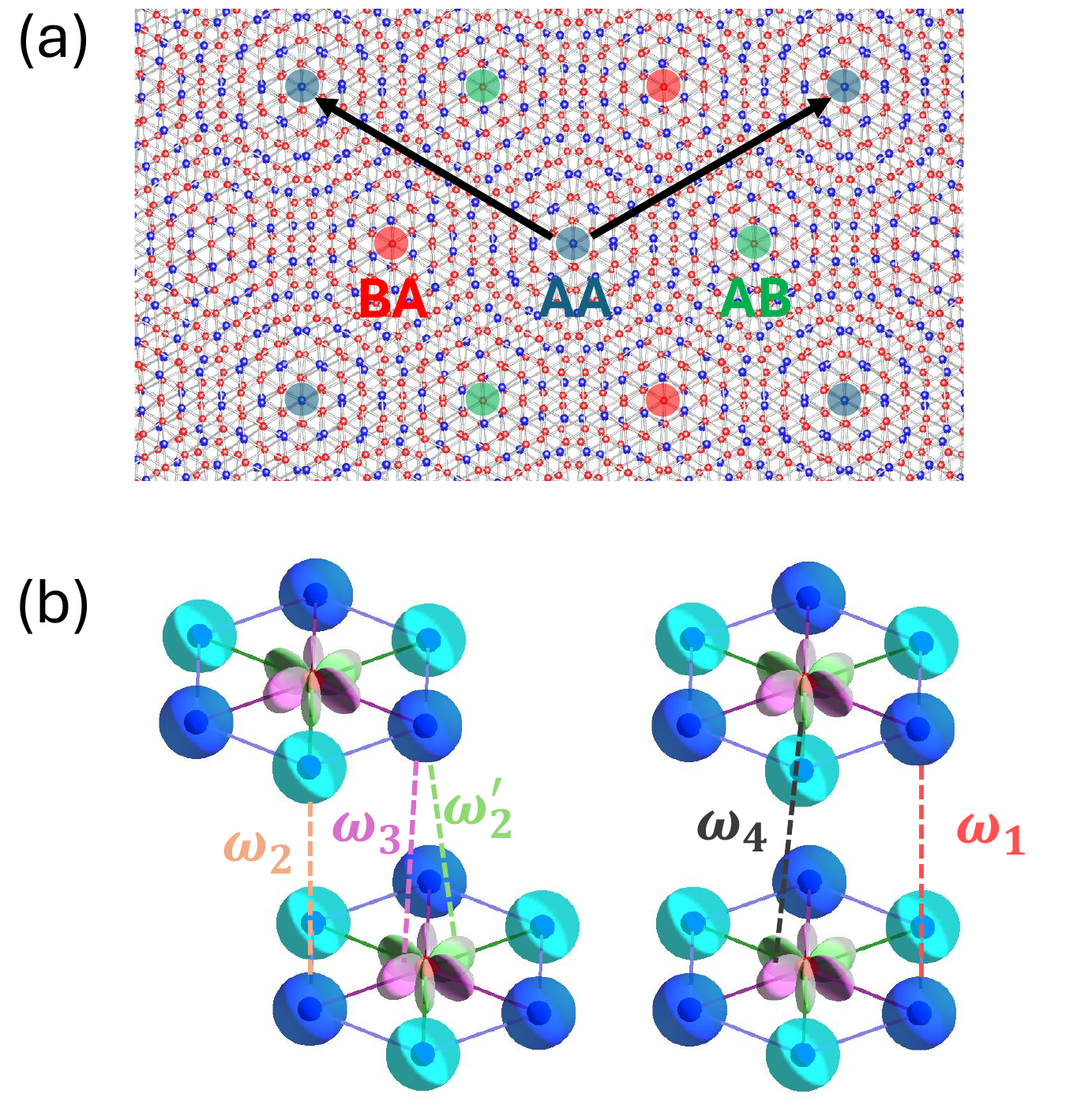}
\caption{ (\textbf{a}) A schematic of the moiré pattern of the twisted bilayer watermill lattice. Colored circles mark the centers of the high-symmetry stacked regions. Black arrows indicate the translation vectors of the moiré unit cell. (\textbf{b}) A schematic for interlayer tunneling in the AB-stacked (left) and AA-stacked (right) regions in the twisted bilayer watermill lattice.}
\label{fig:02}
\end{figure}

\paragraph*{Twisted bilayer watermill lattice (TBWL)---} We investigate moiré physics of TBWL using continuum models constructed via the Bistritzer-MacDonald formalism \cite{PhysRevB.104.115404, doi:10.1073/pnas.1108174108}
\begin{equation}\label{eq:03}
    H(\theta,\xi) = \begin{pmatrix} H_{R}(\frac{\theta}{2},\xi) & T \\ T^{\dagger} & H_{R}(-\frac{\theta}{2},\xi) \end{pmatrix}.
\end{equation}
Here, $\xi$ indicates the monolayer valley degree of freedom, and $\theta$ is the twist angle between the layers. $H_{R}(\theta, \xi)$ denotes the low-energy effective Hamiltonian of the \emph{rotated} monolayer watermill lattice, which can be obtained via a rotation $\theta$ of the Hamiltonian in Eq.~\eqref{eq:02}
\begin{equation}\label{eq:04}
    H_R(\theta,\xi) = v(q_x\hat{x}+\xi q_y\hat{y})\cdot \vec{S}^{(\theta)}.
\end{equation}
Here, $\vec{S}^{(\theta)}\equiv e^{-i\frac{\theta}{2}S_3}\vec{S}e^{i\frac{\theta}{2}S_3}$. Vector $\vec{q}=\vec{k}-K_{\xi}^{(\theta)}$, with $K_{\xi}^{(\theta)}$ is the $K$ point of the $\xi$ valley in the \emph{rotated} monolayer BZ. The interlayer tunneling $T$ takes the matrix form:
\begingroup\makeatletter\def\f@size{7.5}\check@mathfonts
\def\maketag@@@#1{\hbox{\m@th\large\normalfont#1}}
\begin{equation}\label{eq:05}
    T = \begin{pmatrix} \omega_1g(\vec{r}) & \omega_2{'}g(\vec{r}-\vec{r}_{AB}) & \omega_3g(\vec{r}+\vec{r}_{AB}) & \omega_4g(\vec{r}) \\ 
         \omega_2{'}g(\vec{r}+\vec{r}_{AB}) & \omega_1g(\vec{r}) & \omega_2g(\vec{r}-\vec{r}_{AB}) & \omega_3g(\vec{r}+\vec{r}_{AB}) \\ 
         \omega_3g(\vec{r}-\vec{r}_{AB}) & \omega_2g(\vec{r}+\vec{r}_{AB}) & \omega_1g(\vec{r}) & \omega_2{'}g(\vec{r}-\vec{r}_{AB}) \\
         \omega_4g(\vec{r}) & \omega_3g(\vec{r}-\vec{r}_{AB}) & \omega_2{'}g(\vec{r}+\vec{r}_{AB}) & \omega_1g(\vec{r}) \end{pmatrix}.
\end{equation}
\endgroup
The interlayer tunneling matrix element $T_{ij}$ describes tunneling from the $j^{\text{th}}$ orbital in the bottom layer to the $i^{\text{th}}$ orbital in the top layer. Here, $g(\vec{r}) = \sum_i e^{i\vec{q}_i \cdot \vec{r}}$, where $\vec{q}_i$ are momentum vectors connecting valleys from different layers in the BZ. Specifically, $(\vec{q}_1, \vec{q}_2, \vec{q}_3) = \frac{4\pi}{3L_s} (\hat{y}, -\frac{\sqrt{3}}{2}\hat{x} - \frac{1}{2}\hat{y}, \frac{\sqrt{3}}{2}\hat{x} - \frac{1}{2}\hat{y})$. The coordinate origin is set at the center of the AA-stacked region, as illustrated in FIG.~\ref{fig:02}(a), with the positions of different stacked regions given by $\vec{r}_{AB} = -\vec{r}_{BA} = \frac{L_s}{\sqrt{3}}\hat{x}$. Here, $L_s$ is the norm of the translation vector of the moiré cell.

\par Parameters $\omega_i$ represent various interlayer hopping strengths. Specifically, $\omega_1$ gives intra-orbital hopping, $\omega_4$ is intra-sublattice tunneling between $A_1$ and $A_2$ orbitals, $\omega_2$ is inter-sublattice hopping between $B$ and $C$ orbitals, $\omega_2'$ gives inter-orbital hopping between $A_1$-$B$ and $A_2$-$C$ pairs, and $\omega_3$ gives hopping between $A_1$-$C$ and $A_2$-$B$ pairs. Precise values of $\omega_i$ depend on the orbital types within each sublattice. For the TBWL with sublattice symmetry, as an example, we consider a configuration where $A_1$ and $A_2$ orbitals are of $sp_2$ type, while $B$ and $C$ orbitals are of $s$ type, preserving the $C_{3z}$ symmetry (FIG.~\ref{fig:01}). This configuration minimizes the overlap between $A_1$ and $A_2$ orbitals. This also applies to interlayer bonds (FIG.~\ref{fig:02}(b)), so that the hopping parameters exhibit the following hierarchy: $\omega_1 \sim \omega_2 \sim \omega_2' \sim \omega_3 \gg \omega_4$.

\paragraph*{Chiral TBWL---}In the TBWL, we employ a unified hopping scheme for both the inter- and intra-layer interactions, assuming that $\omega_1=\omega_3=\omega_4=0$ and $\omega_2{'}=\frac{\sqrt{3}}{2}\omega_2$. This implies that hoppings within pairs $A_1$-$B$, $A_2$-$C$, and $B$-$C$ occur from the top to the bottom layer in the same way as in the intralayer hopping processes. The preserved symmetries in the monolayer then include  chiral symmetry $\Gamma$, particle-hole symmetry $\Xi$, and the combined symmetry $C_{2z}\mathcal{T}$, are also maintained in the TBWL. These symmetries can be represented by the matrices $\Gamma=\tau_0\otimes\sigma_0\otimes\sigma_3$, $C_{2z}\mathcal{T} = \tau_0\otimes\sigma_1\otimes\sigma_1\mathcal{K}$, and $\Xi=\tau_0\otimes\sigma_0\otimes\sigma_3\mathcal{K}$. Here, $\tau_i$ are Pauli matrices representing the layer degree of freedom. Moreover, the system is controlled by the dimensionless parameter $\alpha\equiv\frac{\omega_2}{vk_\theta}$, where $k_\theta=2k_D\sin(\frac{\theta}{2})$ with $k_D$ being the monolayer Dirac momentum \cite{PhysRevB.104.115404, doi:10.1073/pnas.1108174108, PhysRevB.99.155415}. For our calculations, we set $2vk_D=19.81$ eV. We, therefore, refer to the TBWL in this limit as chiral TBWL (CTBWL) like the chiral TBG (CTBG) \cite{PhysRevLett.122.106405}.

\begin{figure}[h]
\centering
\includegraphics[width=\linewidth]{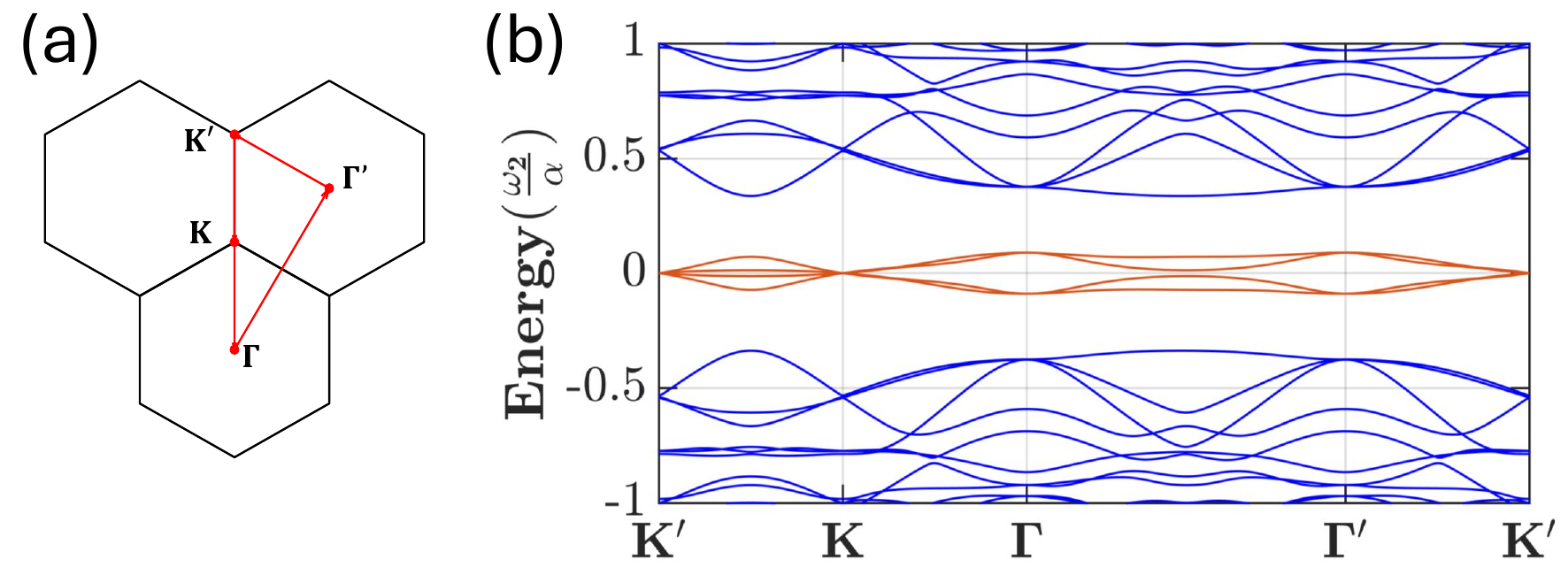}
\caption{ (a) k-path in the moiré Brillouin zone. (b) Twisted bilayer watermill lattice band structure calculated using Eq.~\eqref{eq:03} with $\alpha=0.5$. }
\label{fig:04}
\end{figure}

\begin{figure}[h]
\centering
\includegraphics[width=\linewidth]{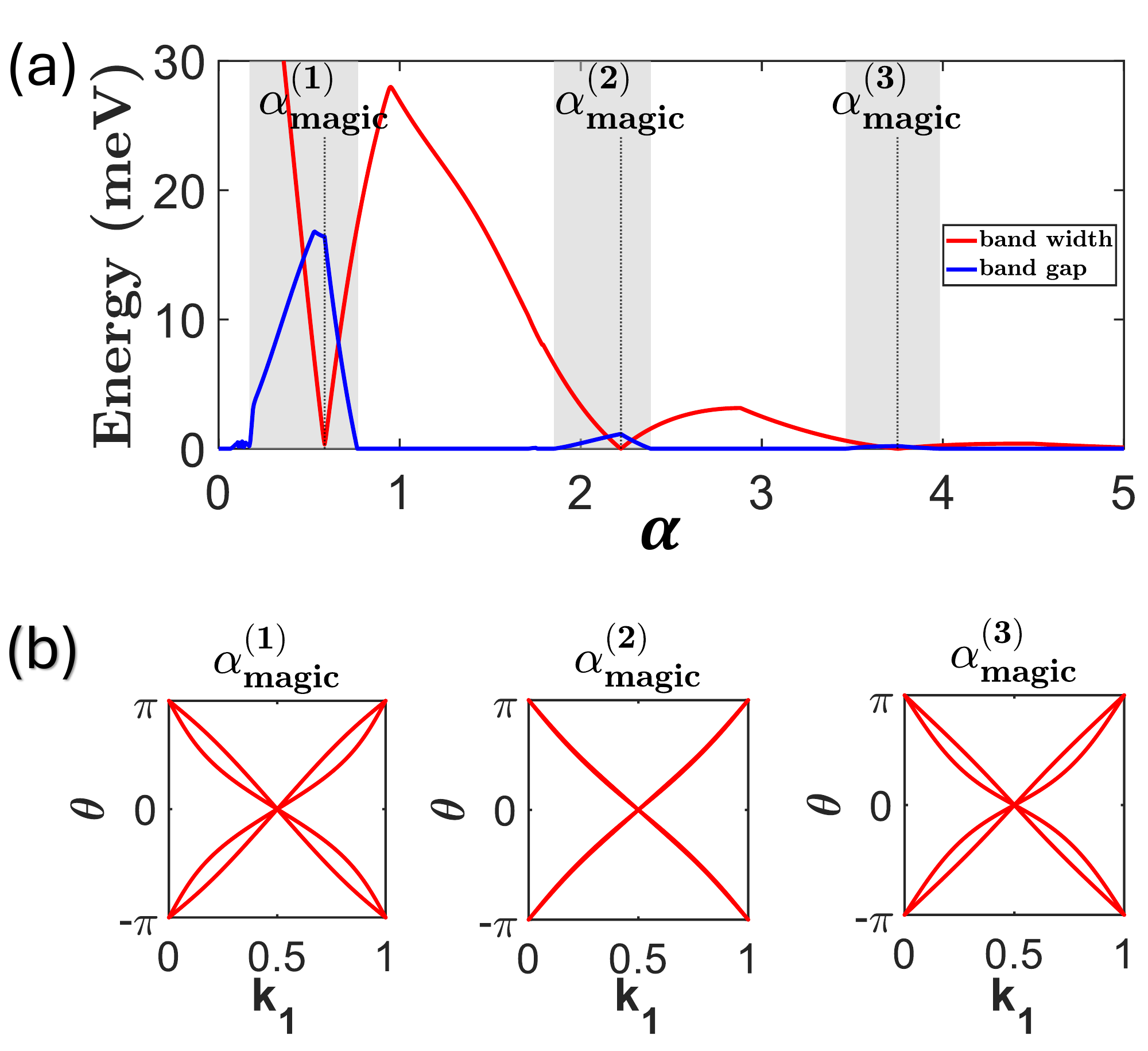}
\caption{ (\textbf{a}) Band gap between the four isolated flat bands near the Fermi level with other high-energy bands (blue) and their bandwidths (red) as a function of $\alpha$. Vertical dashed lines indicate the $i^{\text{th}}$ magic angle, $\alpha_{\text{magic}}^{(i)}$. (\textbf{b}) Wilson loop spectrum of the gapped regions (gray) labeled by the $\alpha_{\text{magic}}^{(i)}$ within.}
\label{fig:05}
\end{figure}

\par The electronic structure of CTBWL now features four isolated flat bands near the Fermi level within specific $\alpha$ ranges (FIG.~\ref{fig:04}). These bands become perfectly flat at certain magic angels, denoted by $\alpha_{\text{magic}}^{(i)}$, with the approximate values of $\alpha_{\text{magic}}^{(1,2,3)} =(0.586, 2.222, 3.749)$ (FIG.~\ref{fig:05}(a)). The emergence of these magic angels can be attributed to chiral symmetry and it can be analytically elucidated through the pseudo-Landau level representation \cite{PhysRevB.99.155415}; see Supplemental Materials (SM) for details \cite{SM}\nocite{ahlfors1979complex, PhysRevB.89.155114, PhysRevB.84.075119, PhysRevB.95.075146, PhysRevLett.107.036601, grosso2013solid}. 

\paragraph*{Quantum geometric effects---} The isolated flat bands in CTBWL have a winding number of $2$ in their Wilson loop spectrum for the first three magic angels (FIG.~\ref{fig:05}(b)). Symmetry $C_{2z}\mathcal{T}$ ensures that the Wilson loop spectrum is symmetric around $0$, as their Wilson loop operator eigenvalues are complex conjugates \cite{PhysRevLett.124.167002, SM}. Such a high winding number further implies an enhanced quantum metric of the flat bands \cite{yu2024universalwilsonloopbound}.

\par The CTBWL exhibits a significantly larger trace of quantum weight $K$ \cite{PhysRevLett.128.087002, PhysRevLett.124.167002, PhysRevB.90.165139, PhysRevX.14.011052} for its isolated flat bands at magic angels compared to CTBG, where $K\cong 2$ at its first magic angle. In contrast, the values are quantified as $K\cong (6.748, 4.844, 4.153)$ for the CTBWL at the magic angels $\alpha_{\text{magic}}^{(1,2,3)}$, respectively. We adopt the continuum model for CTBG in \cite{PhysRevLett.122.106405} and use the same lattice constant and velocity in the calculations for CTBG and CTBWL; see SM for details \cite{SM}. The higher trace of quantum weight is crucial for understanding superconducting properties, as this weight is directly related to the corresponding superfluid weight for weak $s$-wave pairing potentials within the BCS mean-field theory \cite{PhysRevLett.124.167002, PhysRevLett.128.087002, PhysRevB.95.024515, Peotta2015, PhysRevLett.117.045303}.

\par In CTBWL, the superfluid weight is isotropic due to $C_{3z}$ symmetry, allowing for a direct prediction of the BKT transition temperature $T_{\text{BKT}}$ at specific filling factors $\nu$  \cite{JMKosterlitz_1973, PhysRevLett.124.167002}. Based on the calculated trace of the quantum weight $K$ in CTBWL, $T_{\text{BKT}}$ is estimated to be around $0.771\ T_{\text{MF}}$ at its first magic angle $\alpha_{\text{magic}}^{(1)}$ for $\nu = 1/4$. In contrast, the $T_{\text{BKT}}$ of CTBG at its first magic angle with $\nu = 1/4$ is only about $0.468\ T_{\text{MF}}$. At the same electron filling, CTBWL hosts more electrons than CTBG due to its four isolated flat bands per spin per valley, compared to CTBG's two. Moreover, CTBWL exhibits a higher $T_{\text{BKT}}/T_{\text{MF}}$ ratio than CTBG when the number of occupied electrons is fixed at two per moiré cell. In this scenario, CTBWM corresponds to $\nu = 1/8$ and $T_{\text{BKT}} = 0.733\ T_{\text{MF}}$, see SM for details \cite{SM}.

\paragraph{Discussion---}In the watermill lattice, the symmetry $C_{2z}\mathcal{T}$ is crucial for the hopping structure that gives rise to LESs with a pseudospin-3/2 structure. The four-fold degeneracies at valleys in this lattice are resilient to small modifications in hopping strength as long as $C_{3z}$, $C_{2z}\mathcal{T}$ and the chiral symmetry $\Gamma$ are preserved. The hopping strength between orbitals $B$ and $C$, $t_{BC}$, is $\frac{2}{\sqrt{3}}$ times larger than that between other orbitals, $t_{\text{others}}$. However, varying the ratio between $t_{BC}$ and $t_{\text{others}}$ does not lift the four-fold degeneracy but rather modifies the velocities of the Dirac cones, and the pseudospin texture remains a nontrivial winding pattern around the valleys \cite{SM}. Furthermore, magic angels persist in the corresponding CTBWL, see SM and FIG.~S4 for details \cite{SM}. This robustness in hopping strengths should help realize materials with LESs characterized by a pseudospin-3/2 configuration. A promising candidate is the LaAlO$_3$/SrTiO$_3$ (111) quantum well \cite{PhysRevLett.111.126804}. Other interesting possibilities for engineering the watermill lattice would be utilizing covalent-organic frameworks and metal-organic frameworks \cite{doi:10.1021/acs.accounts.0c00652, D1MH00935D, Jiang2021_ACS, Niu2022, https://doi.org/10.1002/cplu.202200359, PhysRevLett.130.036203, Yan2021, Hu2023, Ni2022_1, Ni2022_2, Zhang2023_ACS, Hu2022} or employing CO molecules on Cu(111) surface \cite{Tassi_2024, Gomes2012, Khajetoorians2019}.

\par The CTBWL retains its magic angles even when subjected to a weak onsite potential difference between the $A$ and $B/C$ sites (FIG.~S3(a) \cite{SM}), resulting in the four isolated flat bands splitting into two pairs of entangled flat bands separated by a small band gap as detailed in the SM, see FIG.~S2(c) \cite{SM}. Despite this splitting, the magic angles remain unchanged and the flat bands in each pair continue to exhibit non-trivial windings of their hybrid Wannier centers (FIG.~S3(b) \cite{SM}). Since the onsite potential difference preserves $C_{2z}\mathcal{T}$ symmetry, these non-trivial windings provide a lower bound for the quantum metric of the two isolated flat bands in each group \cite{PhysRevLett.124.167002, yu2024universalwilsonloopbound}. These features should help materials realization of the CTBWL. 

\begin{figure}[h]
\centering
\includegraphics[width=\linewidth]{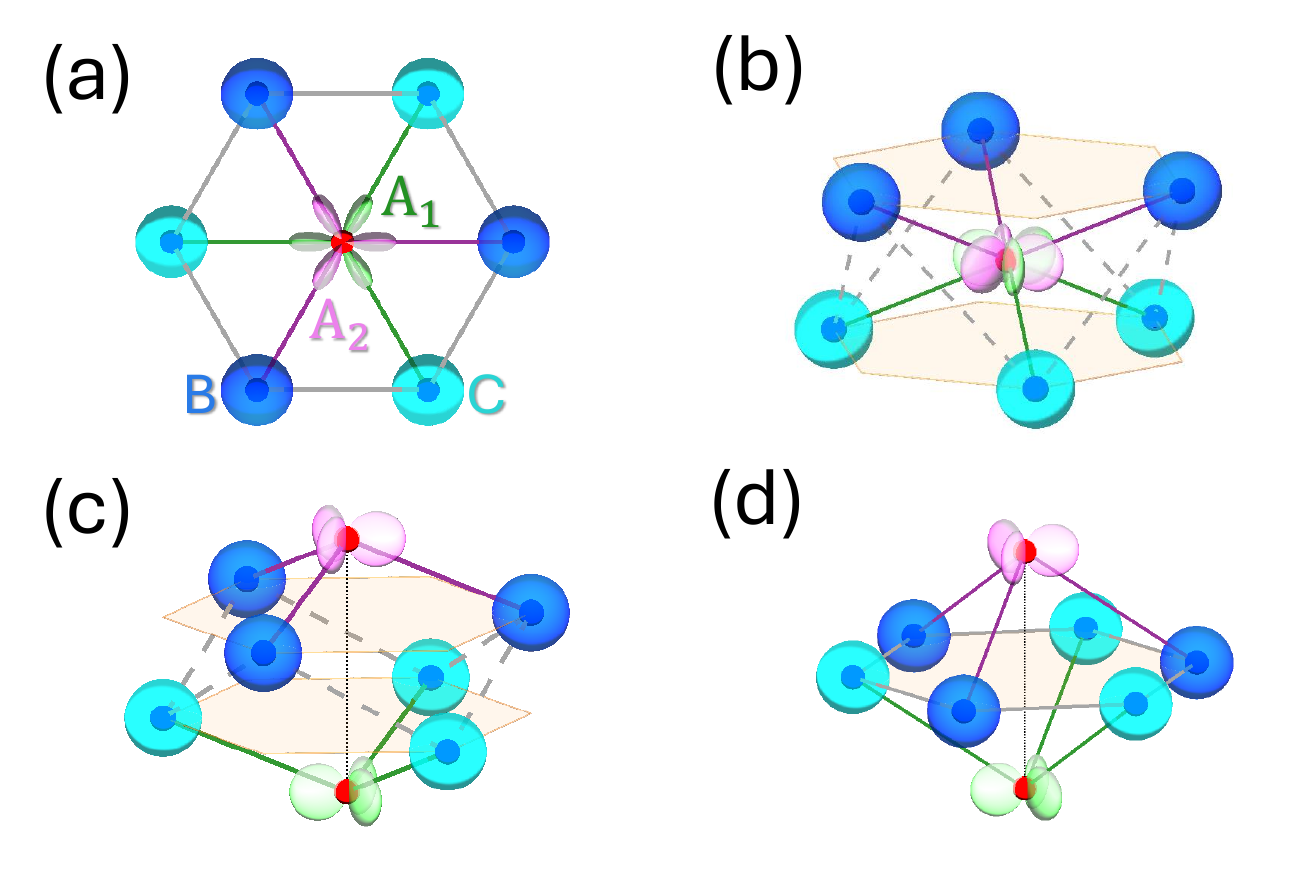}
\caption{ (a) Top view of the modified watermill lattice, where atoms are on different planes but related by the inversion symmetry. (b-d) Side views of the modified watermill lattice with different modifications. Here, orbitals on atoms B and C are assigned distinct colors to emphasize their position differences in these schematic diagrams.}
\label{fig:06}
\end{figure}  

\par We emphasize that the hopping structure of the watermill lattice can also be reproduced in a  \emph{modified watermill lattice}, where the atoms reside in different planes connected by the inversion symmetry $\mathcal{P}$. Here, $\mathcal{PT}$ symmetry substitutes the $C_{2z}\mathcal{T}$ symmetry in the watermill lattice. This configuration could be achieved, for example, by vertically realigning sites $B$ and $C$ with the separation of $A_1$ and $A_2$ orbitals (FIG.~\ref{fig:06}(c)) or by maintaining the $A_1$ and $A_2$ orbitals in place (FIG.~\ref{fig:06}(b)). These modifications should be possible to realize in two-dimensional group IV materials such as Germanene and Stanene \cite{Zhu2015, Zhang2023_germanene, GHADIYALI2018e00341, Broek_2014, PhysRevB.110.165407, https://doi.org/10.1002/smll.201402041, Acun_2015, Davila_2014, XIONG2023115036, Li2014, PhysRevLett.116.256804} as well as MXene \cite{doi:10.1021/acsomega.2c00936, PAPADOPOULOU2022166240, PhysRevB.108.195424, Agapov_2023, D4NR02246G} with B/C and A sites aligning with M and T sites in MXene, respectively. Further studies could seek stable MXene materials or structures with needed orbital arrangements. Separating $A_1$ and $A_2$ orbitals onto distinct, vertically separated sites (FIG.~\ref{fig:06}(d)) could also generate the watermill lattice hopping structure. Generally, these modified structures influence the architecture of interlayer tunneling, thus impacting the moiré physics in TBWL, a topic that extends beyond the scope of this study. Our study provides a new pathway for exploring moiré physics via twisted bilayer high pseudospin fermion lattices.

\paragraph*{Acknowledgement---}
The work at Northeastern University was supported by the National Science Foundation through the Expand-QISE award NSF-OMA-2329067 and benefited from the resources of Northeastern University’s Advanced Scientific Computation Center, the Discovery Cluster, the Massachusetts Technology Collaborative award MTC-22032, and the Quantum Materials and Sensing Institute.


\bibliography{apssamp}
\setcounter{equation}{0}
\setcounter{figure}{0}
\setcounter{table}{0}

\renewcommand{\theequation}{S\arabic{equation}}
\renewcommand{\thefigure}{S\arabic{figure}}
\renewcommand{\thetable}{S\arabic{table}}
\renewcommand{\bibnumfmt}[1]{[S#1]}
\renewcommand{\citenumfont}[1]{S#1}
\newcommand{\bk}{\boldsymbol\kappa}

\newcommand{\beginsupplement}{%
  \setcounter{equation}{0}
  \renewcommand{\theequation}{S\arabic{equation}}%
  \setcounter{table}{0}
  \renewcommand{\thetable}{S\arabic{table}}%
  \setcounter{figure}{0}
  \renewcommand{\thefigure}{S\arabic{figure}}%
  \setcounter{section}{0}
  \renewcommand{\thesection}{S\Roman{section}}%
  \setcounter{subsection}{0}
  \renewcommand{\thesubsection}{S\Roman{section}.\Alph{subsection}}%
}

\clearpage
\pagebreak
\widetext
\begin{center}
\textbf{\large Supplemental Materials: Geometry-Driven Moiré Engineering in Twisted Bilayers of High-Pseudospin Fermions}
\end{center}
\tableofcontents


\section{S1. Origin of the magic angles in chiral twisted bilayer watermill lattice (CTBWL)}
\par The CTBWL refers to the twisted bilayer watermill lattice (TBWL) that involves only the interlayer tunneling between $A_1$-$B$, $A_2$-$C$, and $B$-$C$ pairs, which has the chiral symmetry $\Gamma=\tau_0\otimes\sigma_0\otimes\sigma_3$ as noted in the main text. Here, $\tau_i$ are Pauli matrices representing the degree of freedom of the layer, and $\sigma_i$ are Pauli matrices for the orbital degree of freedom. Then, without loss of generality, we can rewrite Eq.~3 for the $K$ valley of the monolayer BZ in the main text using the basis $\begin{pmatrix}A_1^{(\text{top})}, & A_1^{(\text{bot})}, & C^{(\text{top})}, & C^{(\text{bot})}, & B^{(\text{top})}, & B^{(\text{bot})}, & A_2^{(\text{top})}, & A_2^{(\text{bot})}\end{pmatrix}^T$ to yield \cite{PhysRevLett.122.106405}:
\begin{equation}\label{eq:S1_new}
    H \propto \begin{pmatrix} 0 & 0 & \frac{\sqrt{3}}{2}D(\vec{r}) & 0 \\ 0 & 0 & D^*(-\vec{r}) & \frac{\sqrt{3}}{2}D(\vec{r}) \\ \frac{\sqrt{3}}{2}D^*(-\vec{r}) & D(\vec{r}) & 0 & 0 \\ 0 & \frac{\sqrt{3}}{2}D^*(-\vec{r}) & 0 & 0 \end{pmatrix}.
\end{equation}
Here, $o^{(\text{top/bot})}$ with $o=A_{1(2)}, B, C$ indicates the $o$th orbital on top/bottom layer and,
\begin{equation}\label{eq:S2_new}
    D(\vec{r}) = \begin{pmatrix} -i\bar{\partial} & \alpha t(\vec{r}) \\ \alpha t(-\vec{r}) & -i\bar{\partial} \end{pmatrix},
\end{equation}
where $t(\vec{r})=e^{i\vec{q}_1\cdot\vec{r}} + e^{-i\frac{2\pi}{3}}e^{i\vec{q}_2\cdot\vec{r}} + e^{i\frac{2\pi}{3}}e^{i\vec{q}_3\cdot\vec{r}}$ and $\alpha\equiv\frac{\omega_2}{vk_\theta}$ are as defined in the main text. Note that we have approximate $\vec{S}^{(\theta)}\cong\vec{S}$ since $\theta\ll1$ is assumed. The wave function is decomposed into $\Psi(\vec{r}) = \begin{pmatrix} \psi_{A_1}^{(\vec{k})}(\vec{r}), & \psi_{B}^{(\vec{k})}(\vec{r}), & \psi_{C}^{(\vec{k})}(\vec{r}), & \psi_{A_2}^{(\vec{k})}(\vec{r}) \end{pmatrix}^T$, where $\psi_{o}^{(\vec{k})}(\vec{r})$ with $o=A_{1(2)}, B, C$ are two-component spinors indicating the component of the wave functions on $o$th orbitals in both layers at $\vec{k}$ point in the moiré Brillouin zone (MBZ). Then, the existence of exactly-flat bands at the magic angles implies that:
\begin{equation}\label{eq:S3_new}
    \begin{pmatrix} \frac{\sqrt{3}}{2}D(\vec{r}) & 0 \\ D^*(-\vec{r}) & \frac{\sqrt{3}}{2}D(\vec{r}) \end{pmatrix}\begin{pmatrix} \psi_{C}^{(\vec{k})}(\vec{r}) \\ \psi_{A_2}^{(\vec{k})}(\vec{r}) \end{pmatrix} = 0.
\end{equation}
We anticipate that the flat-band wave functions at magic angles have the form:
\begin{equation}
    \Phi_1^{(\vec{k})}(\vec{r}) = \begin{pmatrix}\phi_{C}^{(\vec{k})}(\vec{r}) \\ 0 \\ 0 \end{pmatrix} + \begin{pmatrix} 0 \\ 0 \\\delta \psi_{C}^{(\vec{k})}(\vec{r}) \end{pmatrix} \text{ and } \Phi_2^{(\vec{k})}(\vec{r}) = \begin{pmatrix} 0 \\ 0 \\ \phi_{A_2}^{(\vec{k})}(\vec{r}) \end{pmatrix},
\end{equation}
where the functions $\phi_{C}^{(\vec{k})}(\vec{r})$ and $\phi_{A_2}^{(\vec{k})}(\vec{r})$ satisfy the equations:
\begin{equation}\label{eq:S5_new}
    D(\vec{r})\phi_{C(A_2)}^{(\vec{k})}(\vec{r}) = 0.
\end{equation}
Given that the wave functions are represented in the basis $e^{i\vec{q}\cdot\vec{r}}$, where $\vec{q}=\vec{k}-K_{\xi}^{(\theta)}$, with $K_{\xi}^{(\theta)}$ as the $K$ point in the $\xi$ valley of the \emph{rotated monolayer} (i.e., top or bottom layer) BZ, there exist two zero modes for Eq.~\eqref{eq:S5_new} at each valley within the MBZ when $\alpha=0$. As the valleys in the MBZ are $C_{3z}$-symmetric locations, the wave functions at these points are defined by the $C_{3z}$ symmetry eigenvalues. Moreover, because the $C_{3z}$ symmetry commutes with the particle-hole symmetry $\Xi$, each zero mode at each valley can be individually analyzed, where the $C_{3z}$ rotation operator for each spinor can be formulated using $\hat{C}_{3z}=\text{diag}(1,e^{-i\frac{2\pi}{3}})e^{\frac{2\pi}{3}\vec{r}\times\nabla}$ resulting from $D(C_{3z}\vec{r})=e^{i\frac{2\pi}{3}}D(\vec{r})$. Since $\alpha\neq0$ does not break the symmetries, the zero modes at each valley persist as $\alpha$ is adiabatically turned on \cite{PhysRevLett.122.106405}.

Since zero-modes always exist at $K, K'$ points in the MBZ for all $\alpha$ and the kinetic part of $D(\vec{r})$ is antiholomorphic in terms of the complex positions $z=x+iy$ and $\bar{z}=x-iy$, we can construct the flat band wave functions at magic angles using:
\begin{align}
    & \Phi_1^{(\vec{k})}(\vec{r}) = h^{(\vec{k})}(z)\begin{pmatrix}\phi_{C}^{(K)}(\vec{r}) \\ 0 \\ 0 \end{pmatrix} + \begin{pmatrix} 0 \\ 0 \\\delta \psi_{C}^{(\vec{k})}(\vec{r}) \end{pmatrix} \label{eq:S6_new}
    \\ \text{and } & \Phi_2^{(\vec{k})}(\vec{r}) = h^{(\vec{k})}(z)\begin{pmatrix} 0 \\ 0 \\ \phi_{A_2}^{(K)}(\vec{r}) \end{pmatrix}, \label{eq:S7_new}
\end{align}
where $h^{(\vec{k})}(z)$ is some holomorphic function \cite{PhysRevLett.122.106405} and $\delta \psi_{C}^{(\vec{k})}$ satisfies the equation:
\begin{equation}\label{eq:S8_new}
    \delta \psi_{C}^{(\vec{k})}(\vec{r}) = -\frac{2}{\sqrt{3}}D^{-1}(\vec{r})D^*(-\vec{r})h^{(\vec{k})}(z)\phi_{C}^{(K)}(\vec{r}).
\end{equation}
Due to Eq.~\eqref{eq:S5_new}, the wave functions $\phi_{C(A_2)}^{(\vec{k})}(\vec{r})$ satisfy the boundary condition:
\begin{equation}
    \phi_{C(A_2)}^{(\vec{k})}(\vec{r}+\vec{a}_{1,2}^{(M)}) = e^{i\vec{k}\cdot\vec{a}_{1,2}^{(M)}}\text{diag}(1,e^{-i\frac{2\pi}{3}})\phi_{C(A_2)}^{(\vec{k})}(\vec{r})
\end{equation}
where $\vec{a}_{1,2}^{(M)}$ are the translation vectors of the moiré cell. On the other hand, due to the $C_{3z}$ symmetry, 
\begin{equation}
    \phi_{C(A_2)}^{(K)}(\vec{r}+\vec{a}_{1,2}^{(M)}) = \text{diag}(1,e^{-i\frac{2\pi}{3}})\phi_{C(A_2)}^{(K)}(\vec{r}).
\end{equation}
Therefore, 
\begin{equation}\label{eq:S10_new}
    h^{(\vec{k})}(z+a_{1,2}^{(M)}) = e^{i\vec{k}\cdot\vec{a}_{1,2}^{(M)}}h^{(\vec{k})}(z),
\end{equation}
where $a_{1,2}^{(M)}\equiv \hat{x}\cdot\vec{a}_{1,2}^{(M)}+i\hat{y}\cdot\vec{a}_{1,2}^{(M)}$. $h^{(\vec{k})}(z)$ must exhibit at least one simple pole. This can be explained as follows: we can always construct a periodic and holomorphic function $g^{(\vec{k})}(z)=e^{-i\text{Re}(\bar{k}z)}h^{(\vec{k})}(z)$ that remains bounded over a compact interval, thanks to its periodicity and holomorphism of $g^{(\vec{k})}(z)$, where $\bar{k}\equiv\hat{x}\cdot\vec{k}-i\hat{y}\cdot\vec{k}$. However, since $h^{(\vec{k})}(z)\neq\alpha e^{i\text{Re}(\bar{k}z)}$ for some constant $\alpha$ in general, $g^{(\vec{k})}(z)$ is not a constant and according to Liouville's theorem \cite{ahlfors1979complex}, it must possess at least one simple pole. Thus, the construction of flat band wave function at magic angles according to Eqs.~\eqref{eq:S6_new} and \eqref{eq:S7_new} is possible only if $\phi_{C(A_2)}^{(K)}(\vec{r}_0)=0$ for some $\vec{r}_0$. When this criterion is combined with Eqs.~\eqref{eq:S5_new} and \eqref{eq:S7_new}, the resulting magic angles (i.e., the values of $\alpha$ that meet all these requirements) can be determined. 

\par Notably, the chirally symmetric flat-band wave functions at magic angles are zero at the center of the AB/BA stacking regions (FIG.~\ref{fig:S1}(b)). Therefore, such a derivation of magic angles in the CTBWL mirrors the one for the chiral twisted bilayer graphene (CTBG), see Ref.~\cite{PhysRevLett.122.106405} for details. CTBWL thus hosts magic angles that are identical to those in CTBG in accord with the numerical findings presented in the main text for CTBWL and the values reported for CTBG in Ref.~\cite{PhysRevLett.122.106405}. Here, we only show the construction of the flat-band wave functions for the $\psi_{C}^{(\vec{k})}(\vec{r}), \psi_{A_2}^{(\vec{k})}(\vec{r})$ components. The other flat-band wave functions can be constructed along similar lines.

\section{S2. Pseudo-Landau level representation of CTBWL at the first magic angle}
\par
\begin{figure}[h]
\centering
\includegraphics[width=\linewidth]{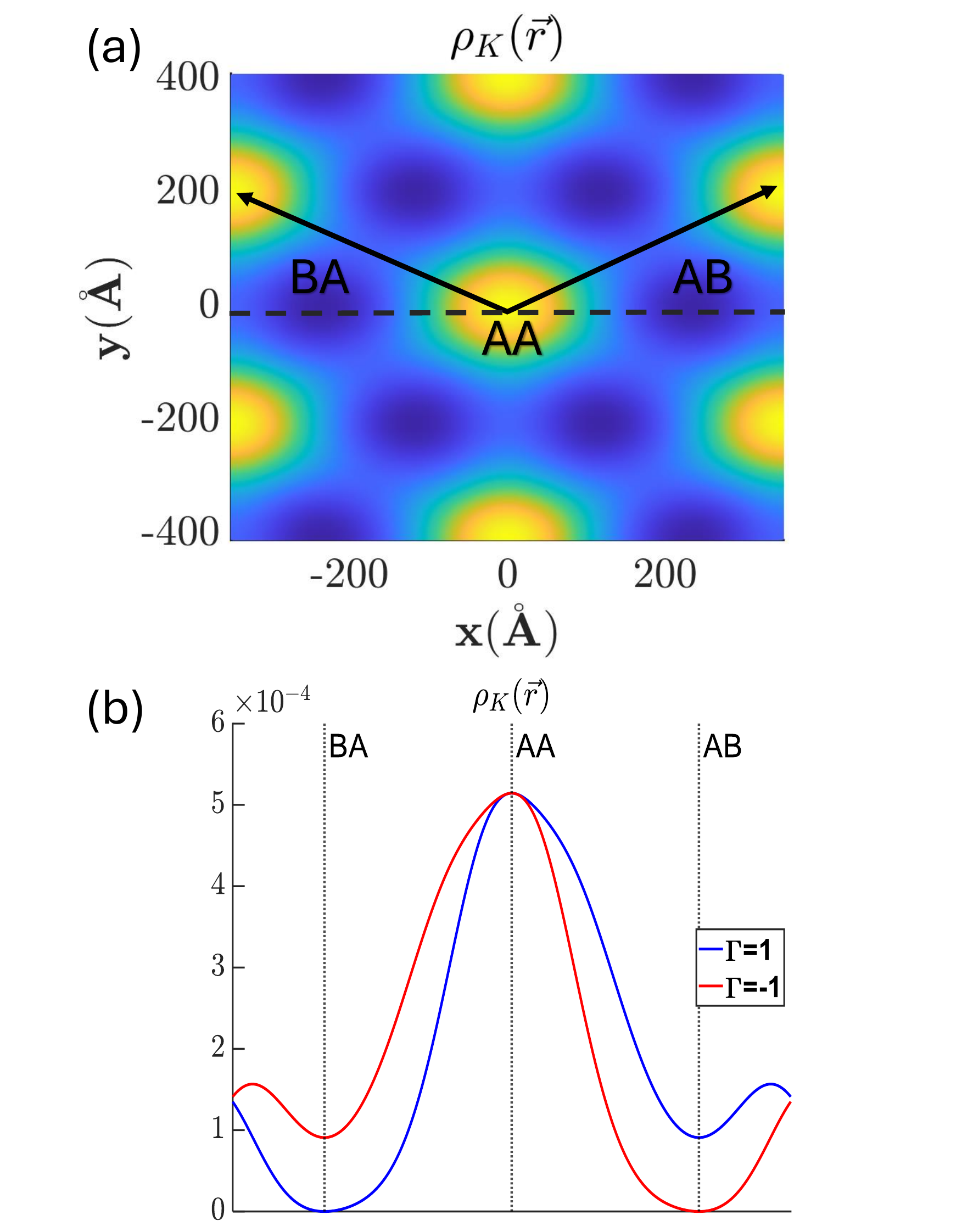}
\caption{ (a) Spatial distribution of the flat-band wave functions at $K$ point in MBZ at the first magic angle for the watermill lattice constant of $a=2.46\text{\AA}$ and twist angle $\theta_{\text{twist}}\cong0.346^\circ$. Black arrows denote the translation vectors of the moiré unit cell. (b) Spatial distribution (up to a normalization factor) of the chirally-symmetric flat-band wave functions at $K$ point in the MBZ with eigenvalues of the chiral symmetry operator $\Gamma=1$ (blue) and $\Gamma=-1$ (red) along the black dashed line in (a) at the first magic angle. }
\label{fig:S1}
\end{figure}

 At the first magic angle $\alpha_{\text{magic}}^{(1)}$, the wave functions of the isolated flat bands are primarily localized around the AA-stacked region ($\vec{r}=0$) as shown in FIG.~\ref{fig:S1}(a). To analyze the Hamiltonian for these flat bands in the limit $\omega_1=\omega_3=\omega_4=0$, we expand the interlayer tunneling of TBWL in Eq.~5 to order $O(\frac{r}{L_s})$ \cite{PhysRevB.99.155415}: 

\begin{equation}\label{eq:15}
    T \cong \omega_2\big\{T_1(1-i\frac{4\pi y}{3L_s})+T_2[1+i\frac{2\pi}{3L_s}(\sqrt{3}x+y)] +T_3[1+i\frac{2\pi}{3L_s}(-\sqrt{3}x+y)]\big\},
\end{equation}
where $T_n = \cos[(n-1)\frac{2\pi}{3}]S_1 - \sin[(n-1)\frac{2\pi}{3}]S_2$. Then, Eq.~\eqref{eq:15} can be interpreted as the effective vector potential in the watermill lattice, given by $\vec{A}=-\frac{2\pi\omega_2}{L_s}(y\hat{x}-\xi x\hat{y})$. The low-energy Hamiltonian in TBWL can, therefore, be expressed as:
\begin{equation}\label{eq:16}
\begin{split}
    H = & (\frac{3t}{2}q_x\tau_0-A_x\tau_2)S_1 + (\frac{3t}{2}\xi q_y\tau_0-A_y\tau_2)S_2
    \\ \rightarrow & (\frac{3t}{2}q_x\tau_0-A_x\tau_3)S_1 + (\frac{3t}{2}\xi q_y\tau_0-A_y\tau_3)S_2,
\end{split}
\end{equation}
where we have applied an appropriate basis transformation on the layer degrees of freedom represented by $\mathbf{\tau}$ in the second line and $\vec{S}^{(\theta)}\cong\vec{S}$ for $\theta\ll1$ is assumed.

\par In the Landau gauge, $\vec{A}=\frac{4\pi\omega_2\xi}{L_s}x\hat{y}$ is the effective vector potential and $B=\frac{4\pi\omega_2\xi}{L_s}$ is the effective field strength for the top layer. By introducing the creation (annihilation) operator for Landau level $a^\dagger(a)\equiv\frac{1}{\sqrt{2\hbar eB}}(\pi_x\pm i\pi_y)$ with the canonical momentum $\vec{\pi}\equiv\vec{k}-e\vec{A}$, as well as the cyclotron frequency $\omega_c=\frac{3t}{2}\sqrt{\frac{2eB}{\hbar}}$, the Hamiltonian in Eq.~\eqref{eq:16} for the top layer can be rewritten as:
\begin{equation}\label{eq:B1}
    H = \hbar\omega_c\begin{pmatrix} 0 & \frac{\sqrt{3}}{2}a^\dagger & 0 & 0 \\ \frac{\sqrt{3}}{2}a & 0 & a^\dagger & 0 \\ 0 & a & 0 & \frac{\sqrt{3}}{2}a^\dagger \\ 0 & 0 & \frac{\sqrt{3}}{2}a & 0 \end{pmatrix}.
\end{equation}
To obtain the Landau levels, we denote the eigenvalue and eigenstate of the $n^{th}$ Landau level as $E_n$ and  $\ket{\psi_n(k)}$, respectively. We assume that $\ket{\psi_n(k)}$can be constructed using the eigenstates of the cyclotron motions, as shown below,
\begin{equation}
\begin{split}
    \ket{\psi_n(k)} = & \frac{e^{ikx}}{\sqrt{2L_xl_B}}\big(u_n\begin{pmatrix} 1 \\ 0 \\ 0 \\ 0 \end{pmatrix}\ket{n} + v_n\begin{pmatrix} 0 \\ 1 \\ 0 \\ 0 \end{pmatrix}\ket{n-1}  + \alpha_n\begin{pmatrix} 0 \\ 0 \\ 1 \\ 0 \end{pmatrix}\ket{n-2} + \beta_n\begin{pmatrix} 0 \\ 0 \\ 0 \\ 1 \end{pmatrix}\ket{n-3}\big).
\end{split}
\end{equation}
Here, $l_B=\sqrt{\frac{3L_s\hbar t}{8\pi\omega_2}}$ is the magnetic length, $L_x$ is the size of the system along the $\hat{x}$ direction, and ($u_n$, $v_n$, $\alpha_n$, $\beta_n$) are the complex coefficients of the cyclotron motion eigenstates. The state $\ket{n}$ satisfies $a\ket{n}=\sqrt{n}\ket{n-1}$ and $a^\dagger\ket{n}=\sqrt{n+1}\ket{n+1}$. From the eigenvalue equation  $H\ket{\psi_n(k)}=E_n\ket{\psi_n(k)}$, we find that $E_n$ satisfies the equation:
\begin{equation}\label{eq:B3}
    \text{det}\begin{pmatrix} E_n & -\hbar\omega_c\frac{\sqrt{3}}{2}\sqrt{n} & 0 & 0 \\ -\hbar\omega_c\frac{\sqrt{3}}{2}\sqrt{n} & E_n & -\hbar\omega_c\sqrt{n-1} & 0 \\ 0 & -\hbar\omega_c\sqrt{n-1} & E_n & -\hbar\omega_c\frac{\sqrt{3}}{2}\sqrt{n-2} \\ 0 & 0 & -\hbar\omega_c\frac{\sqrt{3}}{2}\sqrt{n-2} & E_n \end{pmatrix} = 0.
\end{equation}
Consequently, the solution to Eq.~\eqref{eq:B3} can be obtained as follows:
\begin{align*}
    E_0 = 0, 
    \quad E_1 = \pm\frac{\sqrt{3}}{2}\hbar\omega_c, 
    \quad E_2 = 0 \quad \text{or} \quad \pm\frac{\sqrt{10}}{2}\hbar\omega_c,\text{ and}
    \quad E_n = \pm\frac{\hbar\omega_c}{2}\sqrt{5(n-1)\pm4\sqrt{(n-1)^2+\frac{9}{16}}} \quad \forall n\geq3.
\end{align*}
Similar results can be obtained for the bottom layer. However, the bottom layer experiences the opposite gauge field compared to the top layer, resulting in a minus sign for its vector potential. The Hamiltonian in Eq.~\eqref{eq:B1} for the bottom layer then becomes:

\begin{equation}\label{eq:B4}
    H = \hbar\omega_c\begin{pmatrix} 0 & \frac{\sqrt{3}}{2}a & 0 & 0 \\ \frac{\sqrt{3}}{2}a^\dagger & 0 & a & 0 \\ 0 & a^\dagger & 0 & \frac{\sqrt{3}}{2}a \\ 0 & 0 & \frac{\sqrt{3}}{2}a^\dagger & 0 \end{pmatrix},
\end{equation}
and the ansatz for the eigenstates takes the form:
\begin{equation}
\begin{split}
    \ket{\psi_n(k)} = & \frac{e^{ikx}}{\sqrt{2L_xl_B}}\big(u_n\begin{pmatrix} 1 \\ 0 \\ 0 \\ 0 \end{pmatrix}\ket{n-3} + v_n\begin{pmatrix} 0 \\ 1 \\ 0 \\ 0 \end{pmatrix}\ket{n-2}  + \alpha_n\begin{pmatrix} 0 \\ 0 \\ 1 \\ 0 \end{pmatrix}\ket{n-1} + \beta_n\begin{pmatrix} 0 \\ 0 \\ 0 \\ 1 \end{pmatrix}\ket{n}\big),
\end{split}
\end{equation}
Correspondingly, Eq.~\eqref{eq:B3} becomes:
\begin{equation}\label{eq:B5}
    \text{det}\begin{pmatrix} E_n & -\hbar\omega_c\frac{\sqrt{3}}{2}\sqrt{n-2} & 0 & 0 \\ -\hbar\omega_c\frac{\sqrt{3}}{2}\sqrt{n-2} & E_n & -\hbar\omega_c\sqrt{n-1} & 0 \\ 0 & -\hbar\omega_c\sqrt{n-1} & E_n & -\hbar\omega_c\frac{\sqrt{3}}{2}\sqrt{n} \\ 0 & 0 & -\hbar\omega_c\frac{\sqrt{3}}{2}\sqrt{n} & E_n \end{pmatrix} = 0.
\end{equation}
The Landau levels of the bottom layer are identical to those of the top layer, see  Eq.~\eqref{eq:B3}. Consequently, each layer contributes two zero modes, arising from the states with $n=0$ and $n=2$, resulting in the four isolated flat bands observed at the first magic angle, $\alpha_{\text{magic}}^{(1)}$, in CTBWL.

\section{S3. Wilson-loop winding constrained by the $C_{2z}\mathcal{T}$ symmetry in CTBWL}
\par Following the methods in Ref.~\cite{PhysRevLett.124.167002}, we now show that the hybrid Wannier centers exhibit non-trivial windings in CTBWM, and the $C_{2z}\mathcal{T}$ symmetry ensures that the eigenvalues of the Wilson loop of the isolated flat bands are complex conjugates of each other, that is, their Wilson-loop spectrum is symmetric around zero. More specifically, the plane-wave basis describing the low-energy states of CTBWL can be characterized by the valley degree of freedom $\xi=K/K'$, layer degree of freedom $\tau=\text{top/bottom}$, orbital degree of freedom on each layer $\sigma=A_1^{(\tau)}, B^{(\tau)}, C^{(\tau)}, A_2^{(\tau)}$, reciprocal lattice vector $\vec{Q}$, and the momentum $\vec{k}$, denoted as $\ket{\psi_{\xi,\sigma,\vec{Q},\vec{k}}}$. Note that we have assumed the absence of inter-valley scatterings, ensuring a well-defined valley degree of freedom. The unitary representation of the $C_{2z}\mathcal{T}$ symmetry, denoted as $R(C_{2z}\mathcal{T})$, is defined as:
\begin{equation}
    \begin{cases} 
    (C_{2z}\mathcal{T})\ket{\psi_{\xi,\sigma=A_1^{(\tau)},\vec{Q},\vec{k}}} = R(C_{2z}\mathcal{T})\ket{\psi_{\xi,\sigma=A_2^{(\tau)},\vec{Q},\vec{k}}}
    \\ (C_{2z}\mathcal{T})\ket{\psi_{\xi,\sigma=B^{(\tau)},\vec{Q},\vec{k}}} = R(C_{2z}\mathcal{T})\ket{\psi_{\xi,\sigma=C^{(\tau)},\vec{Q},\vec{k}}}.
    \end{cases}
\end{equation}
Therefore, $R(C_{2z}\mathcal{T})_{\vec{Q},\sigma;\vec{Q}',\sigma'}=\delta_{\vec{Q}\vec{Q}'}(\tau_0\otimes\sigma_1\otimes\sigma_1)_{\sigma\sigma'}$. Note that the complete representation of the $C_{2z}\mathcal{T}$ symmetry $R(C_{2z}\mathcal{T})\textit{K}$ is anti-unitary, where $\textit{K}$ is the anti-linear operator.

\par In our theory, $T^2=1$ since we consider a spinless system. Therefore, we can have a unitary sewing matrix $S(\vec{k})$ for the $C_{2z}\mathcal{T}$ symmetry since there are no Kramers pairs \cite{PhysRevLett.124.167002},
\begin{equation}\label{eq:S0}
    S(\vec{k}) = \Psi^\dagger(\vec{k})R(C_{2z}\mathcal{T})\Psi^*(\vec{k}),
\end{equation}
where $\Psi(\vec{k})=(\ket{\psi_1(\vec{k})},\ket{\psi_2(\vec{k})},\ket{\psi_3(\vec{k})},\ket{\psi_4(\vec{k})})$ with $\ket{\psi_i(\vec{k})}$ be the $i^{\text{th}}$ isolated flat band expanded on the plane-wave basis. Note that $\Psi(\vec{k})\Psi^\dagger(\vec{k})=1$ since $\{\ket{\psi_i(\vec{k})}\}$ is a complete basis set in the subspace spanned by the isolated flat bands. Then, according to Eq.~\eqref{eq:S0}, 
\begin{equation}
    \Psi(\vec{k})S(\vec{k}) = R(C_{2z}\mathcal{T})\Psi^*(\vec{k}).
\end{equation}
The Wilson loop operator for these isolated flat bands is \cite{PhysRevB.89.155114, PhysRevB.84.075119, PhysRevB.95.075146}:
\begin{equation}
    W(k_1) = \lim_{\Delta k\rightarrow0} \Psi^\dagger(k_1,0) \Psi(k_1,\Delta k)\Psi^\dagger(k_1,\Delta k) ... \Psi^\dagger(k_1,2\pi-\Delta k) \Psi(k_1,2\pi),
\end{equation}
of which the complex conjugation is \cite{PhysRevLett.124.167002}:
\begingroup\makeatletter\def\f@size{8}\check@mathfonts
\def\maketag@@@#1{\hbox{\m@th\large\normalfont#1}}
\begin{equation}
\begin{split}
    W^*(k_1) & = \lim_{\Delta k\rightarrow0} \Psi^T(k_1,0) \Psi^*(k_1,\Delta k)\Psi^T(k_1,\Delta k) ... \Psi^T(k_1,2\pi-\Delta k) \Psi^*(k_1,2\pi)
    \\ & = \lim_{\Delta k\rightarrow0} \Psi^T(k_1,0) R^{-1}(C_{2z}\mathcal{T})R(C_{2z}\mathcal{T}) \Psi^*(k_1,\Delta k)\Psi^T(k_1,\Delta k)  R^{-1}(C_{2z}\mathcal{T})R(C_{2z}\mathcal{T}) ... \Psi^T(k_1,2\pi-\Delta k)  R^{-1}(C_{2z}\mathcal{T})R(C_{2z}\mathcal{T}) \Psi^*(k_1,2\pi)
    \\ & = \lim_{\Delta k\rightarrow0} [\Psi(k_1,0)S(k_1,0)]^\dagger \Psi(k_1,\Delta k)S(k_1,\Delta k) [\Psi(k_1,\Delta k)S(k_1,\Delta k)]^\dagger R(C_{2z}\mathcal{T}) ... [\Psi(k_1,2\pi-\Delta k)S(k_1,2\pi-\Delta k)]^\dagger\Psi(k_1,2\pi)S(k_1,2\pi)
    \\ & = S^\dagger(k_1,0)W(k_1)S(k_1,2\pi)
    \\ & = S^\dagger(k_1,0)W(k_1)S(k_1,0).
\end{split}
\end{equation}
\endgroup
Consequently, the eigenvalues of $W(k_1)$ exist either as complex-conjugate pairs or as real values because $W(k_1)$ only differs from its complex conjugate $W^*(k_1)$ by a unitary transformation. Since $W(k_1)=\mathcal{P}e^{-i\oint dk_2' A_2(k_1,k_2')}$, its eigenvalues take the form $e^{i2\pi\theta_i}$, where $\mathcal{P}$ is the path ordering operator, and $A_2(k_1,k_2')=i\Psi^\dagger(k_1,k_2)\partial\Psi(k_1,k_2)/\partial k_2$ represents the non-Abelian Berry connection. The set $\{\theta_i\mod1\}$ is recognized as the spectrum of the Wilson loop \cite{PhysRevLett.107.036601}. Therefore, the Wilson loop exhibits a trivial spectrum if the eigenvalues of $W(k_1)$ are real, i.e., $\theta_i=0,1/2$ for all $k_1$. Conversely, for a Wilson loop with a non-trivial spectrum, its eigenvalues must be complex-conjugate pairs, leading to a Wilson loop spectrum $\{\theta_i\mod1\}$ that is symmetric with respect to $0$ \cite{PhysRevLett.124.167002}.

\section{S4. Numerical computations for the trace of the quantum weight}
\par Trace of the quantum weight \cite{PhysRevX.14.011052} for the bands of interest is:
\begin{equation}\label{eq:S1}
    K = 2\pi\int \frac{d^2k}{(2\pi)^2}\,\big[g_{xx}(\vec{k}) + g_{yy}(\vec{k})\big].
\end{equation}
Here, $g_{\mu\nu}(\vec{k})$ is the quantum metric of the bands of interest, which can be calculated by the projection operator to the bands of interest $P(\vec{k})$ through $g_{\mu\nu}(\vec{k})=\frac{1}{2}\text{Tr}[(\partial_\mu P)(\partial_\nu P)]$ with $\partial_\mu\equiv\partial/\partial{k^\mu}$ and $\text{Tr}\big[ ... \big]$ indicating the trace over quantum states \cite{PhysRevLett.124.167002, PhysRevLett.128.087002}. The expression of $K$ in Eq.~\eqref{eq:S1} can be rewritten into a coordinate-invariant form \cite{PhysRevLett.128.087002}:
\begin{equation}\label{eq:S2}
    K^{(\text{inv})} = 2\pi\int \frac{d^2k}{(2\pi)^2} \sqrt{\det(\eta)}\eta^{\nu\mu}g_{\mu\nu}(\vec{k}) = 2\pi\int \frac{d^2k}{(2\pi)^2} \sqrt{\det(\eta)} \text{tr}\big[\eta^{-1}g(\vec{k})\big],
\end{equation}
where $A_M$ is the area of the moiré unit cell, and $\text{tr}\big[ ... \big]$ indicates the trace over spatial indices. $\eta^{\mu\nu}$ is defined as the inverse of the spatial metric $\eta_{\mu\nu}$ such that $\eta^{\mu\nu}\eta_{\nu\alpha}=\delta^\mu_\alpha$, where $\eta_{\mu\nu}\equiv\vec{b}_\mu\cdot\vec{b}_\nu$ with $\vec{b}_\mu$ being the reciprocal lattice vector. Note that we have used the convention $\vec{a}_i\cdot\vec{b}_j=\delta_{ij}$, where $\vec{a}_i$ is the translation vector. Consider a transformation $\vec{b}_\mu'=\sum_\nu S_{\mu\nu}\vec{b}_\nu$ with $S\in\text{GL}_2(\mathbb{R})$. Then, the quantities $\eta_{\mu\nu}$, $\eta^{\mu\nu}$, and $g_{\mu\nu}$ transform as:
\begin{align*}
    \eta{'} = & S\eta S^T 
    \\ \eta{'}^{-1} = & (S^T)^{-1}\eta^{-1}S^{-1}
    \\ g{'}(\vec{k}) = & Sg(\vec{k})S^T.
\end{align*}
Therefore, $\text{tr}\big[\eta{'}^{-1}g{'}(\vec{k})\big]=\text{tr}\big[\eta^{-1}g(\vec{k})\big]$, and the integarl in Eq.~\eqref{eq:S2} is invariant under transformation of $\{\vec{b}_\mu\}$ with the corresponding Jacobian $\sqrt{\det(\eta)}$ for the integral measure. $K$ in Eq.~\eqref{eq:S1} can thus be recovered by $K^{\text{(inv)}}$ in Eq.~\eqref{eq:S2} by transforming $\{\vec{b}_\mu\}$ into a frame in which $\vec{b}_1\cdot\vec{b}_2=0$ if $\|\vec{b}_1\|=\|\vec{b}_2\|$. For example, if we have $\vec{b}_1=\frac{2\pi}{a}(\frac{1}{2}\hat{x}+\frac{\sqrt{3}}{2}\hat{y})$ and $\vec{b}_2=\frac{2\pi}{a}(-\frac{1}{2}\hat{x}+\frac{\sqrt{3}}{2}\hat{y})$ so $\|\vec{b}_1\|=\|\vec{b}_2\|$, we can transform them into a frame where $\vec{b}_1'=\vec{b}_1-\vec{b}_2=\frac{2\pi}{a}\hat{x}$ and $\vec{b}_2'=\frac{1}{\sqrt{3}}(\vec{b}_1+\vec{b}_2)=\frac{2\pi}{a}\hat{y}$. At that time, $S=\begin{pmatrix} 1 & -1 \\ \frac{1}{\sqrt{3}} & \frac{1}{\sqrt{3}} \end{pmatrix}$, $\vec{b}_1'\cdot\vec{b}_2'=0$, and $\|\vec{b}_1'\|=\|\vec{b}_2'\|$ so $K=K^{(\text{inv})}$.

We further expand Eq.~\eqref{eq:S2} into a more convenient form for numerical computation \cite{PhysRevLett.128.087002}:
\begin{equation}\label{eq:S3}
    K^{(\text{inv})} = 2\pi\int \frac{dk^1dk^2}{(2\pi)^2A_M}\eta^{\nu\mu}\frac{1}{2}\text{Tr}[(\partial_\mu P)(\partial_\nu P)],
\end{equation}
where $A_M$ is the area of the moiré unit cell and $\text{Tr}[ ... ]$ indicates trace over the quantum states. We numerically evaluate $\frac{1}{2}\text{Tr}\big[(\partial_\mu P)(\partial_\nu P)\big]$ using a finite difference method via:
\begin{equation}
\begin{split}
    \frac{1}{2}\text{Tr}\big\{[\partial_\mu P(\vec{k})][\partial_\nu P(\vec{k})]\big\} & \rightarrow \frac{1}{2\epsilon^2}\text{Tr}\big\{[P(\vec{k}+\epsilon\vec{b}_\mu)-P(\vec{k})][P(\vec{k}+\epsilon\vec{b}_\nu)-P(\vec{k})]\big\}
    \\ & = \frac{1}{2\epsilon^2}\text{Tr}\big[P(\vec{k}+\epsilon\vec{b}_\mu)P(\vec{k}+\epsilon\vec{b}_\nu)+P(\vec{k})^2 - P(\vec{k}+\epsilon\vec{b}_\mu)P(\vec{k})-P(\vec{k}+\epsilon\vec{b}_\nu)P(\vec{k})\big]
    \\ & = \frac{1}{2\epsilon^2} \bigg\{\text{Tr}\big[P(\vec{k}+\epsilon\vec{b}_\mu)P(\vec{k}+\epsilon\vec{b}_\nu)\big]+N-\text{Tr}\big[P(\vec{k}+\epsilon\vec{b}_\mu)P(\vec{k})\big]-\text{Tr}\big[P(\vec{k}+\epsilon\vec{b}_\nu)P(\vec{k})\big]\bigg\}
\end{split}
\end{equation}
where $N\equiv\text{Tr}\big[P(\vec{k})\big]$ is the number of bands of interest. Then, by using a $l\times l$ grid on the Brillouin zone for numerical integration, Eq.~\eqref{eq:S3} can be rewritten as:
\begin{equation}\label{eq:S4}
\begin{split}
    K^{(\text{inv})} = \frac{1}{2\pi}\sum_{\vec{k}} \frac{1}{l^2}\frac{\eta^{\mu\nu}}{A_M}\frac{1}{2\epsilon^2} \bigg\{\text{Tr}\big[P(\vec{k}+\epsilon\vec{b}_\mu)P(\vec{k}+\epsilon\vec{b}_\nu)\big]+N-\text{Tr}\big[P(\vec{k}+\epsilon\vec{b}_\mu)P(\vec{k})\big]-\text{Tr}\big[P(\vec{k}+\epsilon\vec{b}_\nu)P(\vec{k})\big]\bigg\},
\end{split}
\end{equation}
which can be computed efficiently by choosing $\epsilon=\frac{1}{l}$. More specifically, we consider the quantity:
\begin{equation}\label{eq:S5}
F_{(1)}(k^1,k^2)\equiv N-\text{Tr}\big[P(k^1+\epsilon,k^2)P(k^1,k^2)\big].
\end{equation}
Note that we have denoted $P(\vec{k}=k^\mu\vec{b}_\mu)$ as $P(k^1,k^2)$. On a $l\times l$ grid of the Brillouin zone with $\epsilon=\frac{1}{l}$, Eq.~\eqref{eq:S5} becomes:
\begin{equation}\label{eq:S6}
    F_{(1)}(n,m) = N-\text{Tr}\big[P(\frac{n+1}{l},\frac{m}{l})P(\frac{n}{l},\frac{m}{l})\big],
\end{equation}
in which $n,m=0, ... ,l-1$. Similarly, we can define: 
\begin{equation}\label{eq:S6.5}
F_{(2)}(k^1,k^2)\equiv N-\text{Tr}\big[P(k^1,k^2+\epsilon)P(k^1,k^2)\big],
\end{equation}
and its corresponding discrete representation: 
\begin{equation}\label{eq:S7}
    F_{(2)}(n,m) \equiv N-\text{Tr}\big[P(\frac{n}{l},\frac{m+1}{l})P(\frac{n}{l},\frac{m}{l})\big].
\end{equation}
Consider another quantity:
\begin{equation}\label{eq:S8}
    \tilde{F}_{(\mu\nu)}(\vec{k}) \equiv \text{Tr}\big[P(\vec{k}+\epsilon\vec{b}_\mu)P(\vec{k}+\epsilon\vec{b}_\nu)\big]-N.
\end{equation}
Then, on a $l\times l$ grid of the Brillouin zone with $\epsilon=\frac{1}{l}$,
\begin{equation}\label{eq:S9}
    \tilde{F}_{(12)}(n,m) = \tilde{F}_{(21)}(n,m) =  \text{Tr}\big[P(\frac{n+1}{l},\frac{m}{l})P(\frac{n}{l},\frac{m+1}{l})\big]-N,
\end{equation}
and: 
\begin{equation}\label{eq:S10}
    \tilde{F}_{(11)}(n,m) = \text{Tr}\big[P(\frac{n+1}{l},\frac{m}{l})P(\frac{n+1}{l},\frac{m}{l})\big]-N = 0.
\end{equation}
Similarly, $\tilde{F}_{(22)}(n,m)=0$. Note that $P(1,\frac{m}{l})=V(\vec{b}_1)P(0,\frac{m}{l})V^{\dagger}(\vec{b}_1)$ with $m\neq l$ and $P(\frac{n}{l},1)=V(\vec{b}_2)P(\frac{n}{l},0)V^{\dagger}(\vec{b}_2)$ with $n\neq l$, where $V(\vec{b}_\mu)$ is the embedding matrix in the direction of the reciprocal lattice vector $\vec{b}_\mu$ such that $(V(\vec{b}_\mu))_{ij}=\delta_{ij}e^{-i2\pi\vec{b}_\mu\cdot r_i}$ \cite{PhysRevLett.128.087002}. Therefore, combining the results from Eq.~\eqref{eq:S5} to Eq.~\eqref{eq:S10}, Eq.~\eqref{eq:S4} can be computed efficiently through:
\begin{equation}
    K^{(\text{inv})} = \frac{1}{2\pi A_M}\sum_{n,m=0}^{l-1} \bigg[ \eta^{12}\tilde{F}_{(12)}(n,m) + \sum_{\mu=1}^2 (\eta^{\mu\mu}+\eta^{12})F_{(\mu)}(n,m) \bigg].
\end{equation}

\section{S5. Geometric contribution to the superfluid weight in CTBWL}
We consider a mean-field theory described by a Bogoliubov-de-Gennes Hamiltonian with a pairing potential \cite{PhysRevLett.124.167002}
\begin{equation}\label{eq:06}
\Delta(\vec{k}) = [\Delta(T)P(\vec{k})]U.
\end{equation}
Here, $\Delta(T)\in\mathbb{R}$, $P(\vec{k})$ is the projection operator to the flat bands, and $U$ is the unitary part of the time-reversal operator. We use the fact that the flat bands are exactly flat with a substantial band gap $E_{\text{gap}}$ between them and other high-energy bands at magic angles, such that there is no pairing between the flat bands and other high-energy bands. Also, $E_{\text{gap}}\gg\Delta(T)$ is assumed since the pairing potentials are weak. Then, the geometric contribution to their superfluid weight dominates \cite{PhysRevB.95.024515, Peotta2015} and the resulting isotropic superfluid weight is give by: 
\begin{equation}\label{eq:07}
D_s(T) = \frac{\delta^{\mu\nu}D_{\mu\nu}(T)}{2} = \frac{e^2\Delta(T)^2}{4\pi\hbar^2\sqrt{(E_0-\mu)^2+\Delta(T)^2}}\tanh({\frac{\beta\sqrt{(E_0-\mu)^2+\Delta(T)^2}}{2}})K
\end{equation}
for the isolated flat bands in CTBWL at a magic angle, where the chemical potential, $\mu$, is related through the filling factor, $\nu$, by:
\begin{equation}\label{eq:07.5}
    \nu = \frac{1}{2}(1-\frac{E_0-\mu}{\sqrt{(E_0-\mu)^2+\Delta(T)^2}}) + \frac{1}{e^{\beta\sqrt{(E_0-\mu)^2+\Delta(T)^2}}+1},
\end{equation}
along the lines of  Ref. \onlinecite{PhysRevLett.124.167002}. Here, $E_0$ is the flat band energy and $\beta\equiv (k_BT)^{-1}$. Further, we assume the celebrated scaling form of the order parameter $\Delta(T)\cong3k_BT_{\text{MF}}\sqrt{1-\frac{T}{T_{\text{MF}}}}$ around the mean-field transition temperature, $T_{\text{MF}}$, in the BCS theory \cite{grosso2013solid}. Berezinskii-Kosterlitz-Thouless (BKT) transition temperature, $T_{\text{BKT}}$, is then given by\cite{JMKosterlitz_1973},  
\begin{equation}\label{eq:08}
k_BT_{\text{BKT}} = \frac{\pi\hbar^2}{8e^2}D_s(T_{\text{BKT}}).
\end{equation}
The BKT transition temperature can be obtained by self-consistently solving Eq.~\eqref{eq:08} with Eq.~\eqref{eq:07}, where the chemical potential is obtained by solving Eq.~\eqref{eq:07.5} for a given filling factor, $\nu$. The trace of quantum weight originating from every spin and valley was included in obtaining the numerical values listed in the main text. 

\section{S6. Modified hopping strength in watermill lattice and CTBWL}
\par
\begin{figure}[h]
\centering
\includegraphics[width=\linewidth]{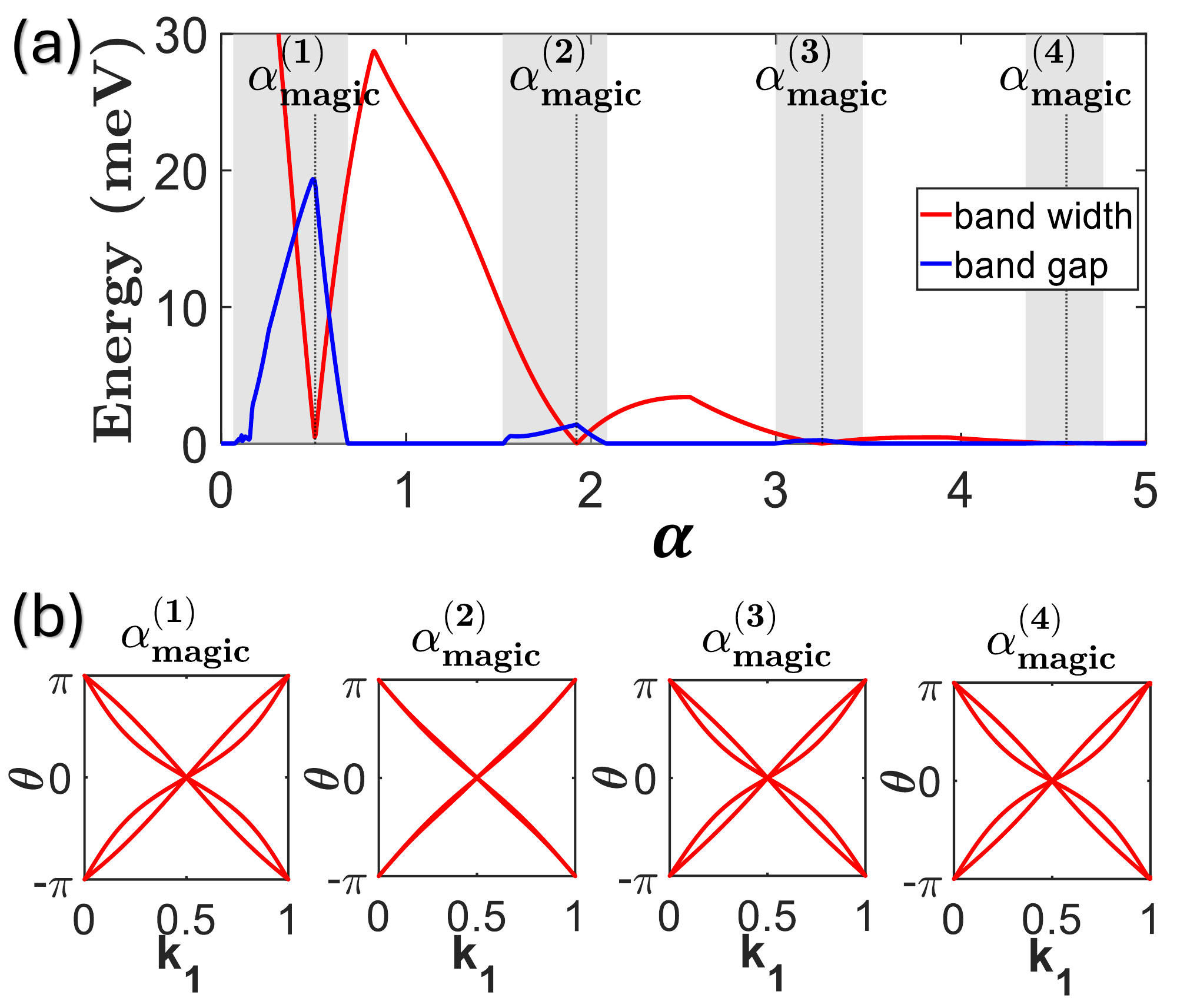}
\caption{ (\textbf{a}) Band gap between the four isolated flat bands near the Fermi level with other high-energy bands (blue) and their bandwidth (red) as a function of $\alpha$ in CTBWL with uniform interlayer tunneling $\omega_2{'}=\omega_2$. Vertical dashed lines indicate the $i^{\text{th}}$ magic angle as $\alpha_{\text{magic}}^{(i)}$. The magic angles are: $\alpha_{\text{magic}}^{(1)}\cong0.508$, $\alpha_{\text{magic}}^{(2)}\cong1.921$, $\alpha_{\text{magic}}^{(3)}\cong3.25$, and $\alpha_{\text{magic}}^{(4)}\cong4.57$. (\textbf{b}) Wilson-loop spectrum of the gapped regions (gray) labeled by the $\alpha_{\text{magic}}^{(i)}$ within.}
\label{fig:S4}
\end{figure}

 If the hopping strength between orbitals $B$ and $C$ has a different ratio $\gamma\in\mathbb{R}$ than the strength of hopping between other orbitals, Eq.~\eqref{eq:02} can be written as:
\begin{equation}\label{eq:St1}
    H_0'(\vec{k}) = t\begin{pmatrix} 0 & \frac{\sqrt{3}}{2}k_- & 0 & 0 \\ \frac{\sqrt{3}}{2}k_+ & 0 & \gamma k_- & 0 \\ 0 & \gamma k_+ & 0 & \frac{\sqrt{3}}{2}k_- \\ 0 & 0 & \frac{\sqrt{3}}{2}k_+ & 0 \end{pmatrix}.
\end{equation}
Here, without loss of generality, we consider the low-energy states at $K$-valley by taking $\xi=+$. Then, the eigenenergies of Eq.~\eqref{eq:St1} are:
\begin{equation}\label{eq:St2}
E=\zeta t\frac{\|\vec{k}\|}{2}\sqrt{(3+2\gamma^2)+2\lambda|\gamma|\sqrt{3+\gamma^2}}
\end{equation}
with the corresponding eigenstates labeled as $\ket{\zeta,\lambda}$, where $\zeta,\lambda=\pm1$. Equation~\eqref{eq:St2} reduces to the dispersion of Eq.~\eqref{eq:02} when $|\gamma|=1$. Since $|3+2\gamma^2| > |2\gamma\sqrt{3+\gamma^2}|$, the dispersion only becomes degenerate at $\|\vec{k}\| = 0$ with a four-fold degeneracy. Note that Eq.~\eqref{eq:St1} still preserves the $C_{2z}\mathcal{T}$ symmetry and the chiral symmetry $\Gamma$ for arbitrary value of $\gamma$. The pseudospin texture for each band is: 
\begin{equation}
\bra{\zeta,\lambda}\vec{S}(\vec{k})\ket{\zeta,\lambda}=\zeta\frac{2\vec{k}}{3\|\vec{k}\|}\sqrt{3+2\gamma^2+2\lambda|\gamma|\sqrt{3+\gamma^2}}\bigg[3+\gamma+\lambda\,\text{sign}(\gamma)\sqrt{3+\gamma^2}\bigg].
\end{equation}
Since the pseudospin texture $\vec{S}(\vec{k})\propto\zeta\frac{\vec{k}}{\|\vec{k}\|}$, each band has a non-trivial winding pattern of the pseudospin texture with the winding number $w=\zeta$. Similarly, the bands of the low-energy states at the $K'$-valley would exhibit a pseudospin texture $\vec{S}(\vec{k})\propto-\zeta\frac{\vec{k}}{\|\vec{k}\|}$ with the winding number $w = -\zeta$. However, the magic angles in the corresponding CTBWL persist at the same values, and Eq.~\eqref{eq:S1_new} becomes:
\begin{equation}\label{eq:S45_new}
    H = \begin{pmatrix} 0 & 0 & \frac{\sqrt{3}}{2}D(\vec{r}) & 0 \\ 0 & 0 & \tilde{D}^*(-\vec{r}) & \frac{\sqrt{3}}{2}D(\vec{r}) \\ \frac{\sqrt{3}}{2}D^*(-\vec{r}) & \tilde{D}(\vec{r}) & 0 & 0 \\ 0 & \frac{\sqrt{3}}{2}D^*(-\vec{r}) & 0 & 0 \end{pmatrix},
\end{equation}
where 
\begin{equation}
    \tilde{D}(\vec{r}) = \begin{pmatrix} -i\gamma \bar{\partial} & \alpha t(\vec{r}) \\ \alpha t(-\vec{r}) & -i\gamma \bar{\partial} \end{pmatrix}.
\end{equation}
The flat-band wave function in Eq.~\eqref{eq:S7_new} thus remains the same but the one in Eqs.~\eqref{eq:S6_new} and \eqref{eq:S8_new} would be modified:
\begin{equation}
    \delta \psi_{C}^{(\vec{k})}(\vec{r}) = -\frac{2}{\sqrt{3}}D^{-1}(\vec{r})\tilde{D}^*(-\vec{r})h^{(\vec{k})}(z)\phi_{C}^{(K)}(\vec{r}).
\end{equation}
Nonetheless, based on the results of the preceding SM section, the method for finding magic angles (Eq.~\eqref{eq:S7_new}) stays unchanged, and the magic angles remain consistent even if the interlayer tunneling strength between orbitals $B$ and $C$ is altered to be $\delta\cdot\alpha$ for a certain constant value $\delta$. Generally speaking, as long as the structure of $D(\vec{r})$ in Eq.~\eqref{eq:S45_new} remains intact, the magic angles retain the values they have in the CTBWL with unmodified hopping strengths. In any event, the magic angles can still be calculated using the techniques already discussed above. An illustrative example is presented in FIG.~\ref{fig:S4}, where $\omega_2'=\omega_2=\alpha vk_\theta$ so that Eq.~\eqref{eq:S45_new} becomes:
\begin{equation}
    H = \begin{pmatrix} 0 & 0 & \frac{\sqrt{3}}{2}\Delta(\vec{r}) & 0 \\ 0 & 0 & D^*(-\vec{r}) & \frac{\sqrt{3}}{2}\Delta(\vec{r}) \\ \frac{\sqrt{3}}{2}\Delta^*(-\vec{r}) & D(\vec{r}) & 0 & 0 \\ 0 & \frac{\sqrt{3}}{2}\Delta^*(-\vec{r}) & 0 & 0 \end{pmatrix}.
\end{equation}
Here:
\begin{equation}
    \Delta(\vec{r}) = \begin{pmatrix} -i\bar{\partial} & \tilde{\alpha} t(\vec{r}) \\ \tilde{\alpha} t(-\vec{r}) & -i\bar{\partial} \end{pmatrix},
\end{equation}
where $\tilde{\alpha}=\frac{2}{\sqrt{3}}\alpha$. As a result, the magic angles are roughly $\frac{\sqrt{3}}{2}$ times their original values, which is consistent with the numerical results shown in FIG.~\ref{fig:S4}.

\par Notably, the pseudo-Landau level representation for the corresponding CTBWL at magic angles continues to produce four zero modes. As previously illustrated, these distinct flat bands can be interpreted as zero modes of the CTBWL. By modifying the hopping ratio, the method described in Eq.~\eqref{eq:B3} can be applied to calculate the pseudo-Landau levels' energy, yielding the following values for the two layer:
\begin{align*}
    & E_0 = 0, 
    \quad E_1 = \pm\frac{\sqrt{3}}{2}\hbar\omega_c, 
    \quad E_2 = 0 \quad \text{or }\pm\frac{\sqrt{6+4\gamma^2}}{2}\hbar\omega_c,
    \\ & \text{and } E_n = \pm\frac{\hbar\omega_c}{2}\sqrt{(3+2\gamma^2)(n-1)\pm\sqrt{4\gamma^2(3+\gamma^2)(n-1)^2+9}} \quad \forall n\geq3,
\end{align*}
demonstrating that the zero modes still originate from the $n=0$ and $n=2$ pseudo-Landau levels.

\section{S7. Effect of onsite potential difference in CTBWL}
\par
\begin{figure}[h]
\centering
\includegraphics[width=\linewidth]{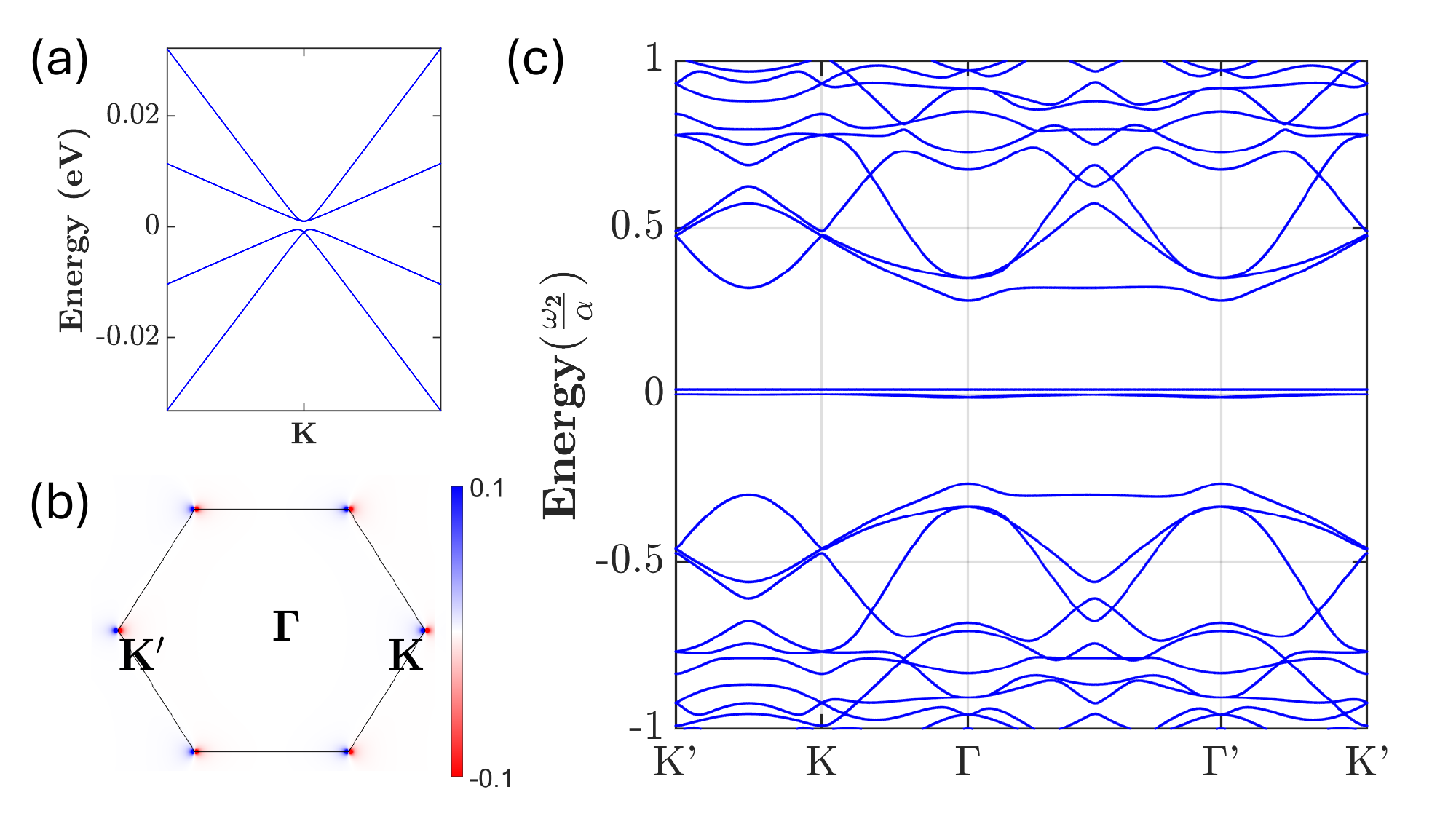}
\caption{ (\textbf{a}) Band structure of the watermill lattice around $K$ valley with onsite energy difference $\Delta E = 1$ meV between site $A$ and $B/C$ sites, and (\textbf{b}) the corresponding valence band Berry curvature distribution in the monolayer Brillouin zone. (\textbf{c}) Band structure of the corresponding CTBWL at the first magic angle. }
\label{fig:S2}
\end{figure}

 The onsite potentials break the sublattice symmetry in the watermill lattice and CTBWL, opening the gap at the $K,K'$ valley. We now consider the effects of onsite potential difference between the sites $A$ and $B/C$. In the watermill lattice, the Berry curvature accumulates around the valley after gap-opening due to differences in onsite potentials. The band structure still hosts a two-fold degeneracy in both the valence and conduction bands. The onsite potential induces a Dirac cone and a quadratic touching in the valence and conduction bands, respectively, with zero total Berry curvature around the valley (FIGS.~\ref{fig:S2}(a) and ~\ref{fig:S2}(b)).

\par
\begin{figure}[h]
\centering
\includegraphics[width=\linewidth]{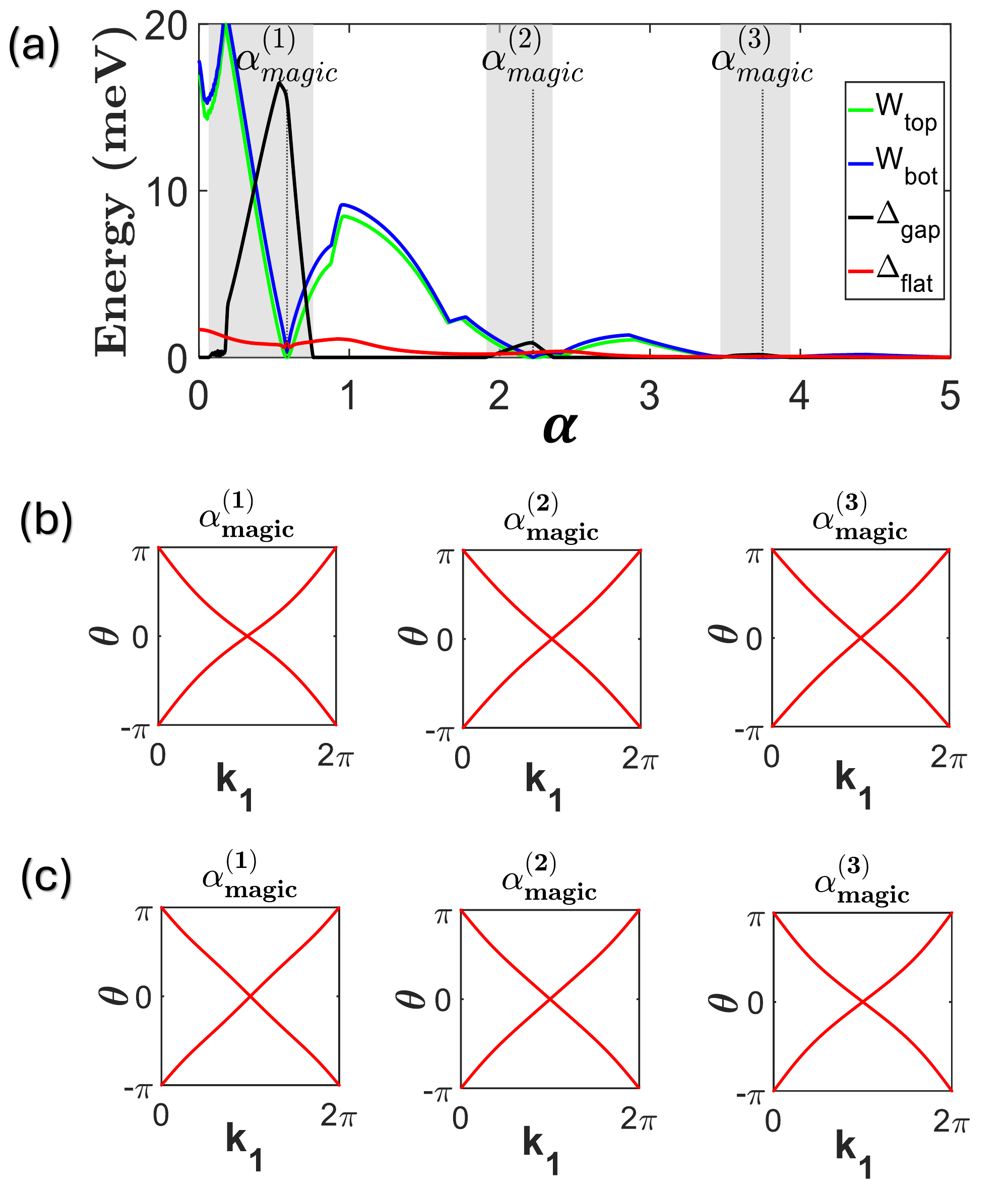}
\caption{ (\textbf{a}) The smallest band gap between the four isolated flat bands near the Fermi level with other high-energy bands ($\Delta_{\text{gap}}$; black). Gap between the two groups of flat bands ($\Delta_{\text{flat}}$; red), and the bandwidth of the flat bands in each group ($W_{\text{top}}$ and $W_{\text{bot}}$; blue and green, respectively) as a function of $\alpha$ in CTBWL subjected to onsite potential difference $\Delta E=1$ meV between the sites A and B/C. The indices top and bot indicate the top and bottom pairs of flat bands. Vertical dashed lines indicate the $i^{\text{th}}$ magic angle, $\alpha_{\text{magic}}^{(i)}$. (\textbf{b}) Wilson-loop spectrum of the top two flat bands in the gapped regions (gray) labeled by the $\alpha_{\text{magic}}^{(i)}$ within. (\textbf{c}) Wilson-loop spectrum of the bottom two flat bands in the gapped regions (gray) labeled by the $\alpha_{\text{magic}}^{(i)}$ within.}
\label{fig:S3}
\end{figure}

 In the pseudo-Landau level representation for the first magic angle in CTBWL subjected to different onsite potentials on various orbitals, Eq.~\eqref{eq:B3} (i.e., pseudo-Landau levels at the top layer) becomes:
\begin{equation}\label{eq:C1}
    \text{det}\begin{pmatrix} E_n^{(\text{top})}-\Delta E_{A_1} & -\hbar\omega_c\frac{\sqrt{3}}{2}\sqrt{n} & 0 & 0 \\ -\hbar\omega_c\frac{\sqrt{3}}{2}\sqrt{n} & E_n^{(\text{top})}-\Delta E_B & -\hbar\omega_c\sqrt{n-1} & 0 \\ 0 & -\hbar\omega_c\sqrt{n-1} & E_n^{(\text{top})}-\Delta E_C & -\hbar\omega_c\frac{\sqrt{3}}{2}\sqrt{n-2} \\ 0 & 0 & -\hbar\omega_c\frac{\sqrt{3}}{2}\sqrt{n-2} & E_n^{(\text{top})}-\Delta E_{A_2} \end{pmatrix} = 0.
\end{equation}
The solution of Eq.~\eqref{eq:B4} is straightforwardly shown to be: $E_0^{(\text{top})}=\Delta E_{A_1}$ and $E_n^{(\text{top})}=E_n^{(\text{top})}(\hbar\omega_c,\Delta E_i)\neq \Delta E_i \text{ or } 0$ for $n\neq0$. Similarly, Eq.~\eqref{eq:B5} (i.e., pseudo-Landau levels at the bottom layer) becomes:
\begin{equation}\label{eq:C2}
    \text{det}\begin{pmatrix} E_n^{(\text{bot})}-\Delta E_{A_1} & -\hbar\omega_c\frac{\sqrt{3}}{2}\sqrt{n-2} & 0 & 0 \\ -\hbar\omega_c\frac{\sqrt{3}}{2}\sqrt{n-2} & E_n^{(\text{bot})}-\Delta E_B & -\hbar\omega_c\sqrt{n-1} & 0 \\ 0 & -\hbar\omega_c\sqrt{n-1} & E_n^{(\text{bot})}-\Delta E_C & -\hbar\omega_c\frac{\sqrt{3}}{2}\sqrt{n} \\ 0 & 0 & -\hbar\omega_c\frac{\sqrt{3}}{2}\sqrt{n} & E_n^{(\text{bot})}-\Delta E_{A_2}\end{pmatrix} = 0,
\end{equation}
which has the solution for $E_0^{(\text{bot})}=\Delta E_{A_2}$ and $E_n^{(\text{bot})}=E_n^{(\text{bot})}(\hbar\omega_c,\Delta E_i)\neq \Delta E_i \text{ or } 0$ for $n\neq0$. In our case, the CTBWL subject to an onsite potential difference $\Delta E$ between the sites A and B/C such that $\Delta E_{A_1}=\Delta E_{A_2}=\Delta E$ and $\Delta E_{B}=\Delta E_{C}=-\Delta E$, and Eqs.~\eqref{eq:C1} and \eqref{eq:C2} has the same solution. When $\Delta E\ll1$, the value of $E_2^{(\text{top/bot})}$ that is closest to 0 can be approximated by $E_2^{(\text{top/bot})}\cong-\frac{\Delta E}{5}$. Therefore, together with the two flat bands from the solution $E_0^{(\text{top/bot})}=\Delta E$ in both layers, CTBWL still hosts four isolated flat bands near the Fermi level subjected to a weak onsite potential difference between A and B/C sites. The isolated flat bands thus split into two groups separated by a small band gap with two entangled flat bands in each group, which is confirmed by our numerical results, see FIG.~\ref{fig:S2}(c). The flat bands in the group with higher energy are exactly flat, while the flat bands in the group with lower energy have a very small but finite bandwidth at the original magic angles (FIG.~\ref{fig:S3}(a)). The flat bands in each group still host non-trivial windings of their hybrid Wannier centers (FIGS.~\ref{fig:S3}(b) and (c)). Since the considered onsite potential difference preserves $C_{2z}\mathcal{T}$ symmetry, such a non-trivial winding of hybrid Wannier centers provides a lower bound for the trace of quantum metric for the two isolated flat bands in each group \cite{PhysRevLett.124.167002}.

\newpage

\newpage

\newpage
\enlargethispage{\baselineskip}

\newpage

\newpage
\end{document}